\PassOptionsToPackage{unicode}{hyperref}
\PassOptionsToPackage{hyphens}{url}
\documentclass[
]{article}
\usepackage{amsmath,amssymb}
\usepackage{lmodern}
\usepackage{iftex}
\ifPDFTeX
  \usepackage[T1]{fontenc}
  \usepackage[utf8]{inputenc}
  \usepackage{textcomp} 
\else 
  \usepackage{unicode-math}
  \defaultfontfeatures{Scale=MatchLowercase}
  \defaultfontfeatures[\rmfamily]{Ligatures=TeX,Scale=1}
\fi
\IfFileExists{upquote.sty}{\usepackage{upquote}}{}
\IfFileExists{microtype.sty}{
  \usepackage[]{microtype}
  \UseMicrotypeSet[protrusion]{basicmath} 
}{}
\makeatletter
\@ifundefined{KOMAClassName}{
  \IfFileExists{parskip.sty}{%
    \usepackage{parskip}
  }{
    \setlength{\parindent}{0pt}
    \setlength{\parskip}{6pt plus 2pt minus 1pt}}
}{
  \KOMAoptions{parskip=half}}
\makeatother
\usepackage{xcolor}
\usepackage[margin=1in]{geometry}
\usepackage{graphicx}
\makeatletter
\def\maxwidth{\ifdim\Gin@nat@width>\linewidth\linewidth\else\Gin@nat@width\fi}
\def\maxheight{\ifdim\Gin@nat@height>\textheight\textheight\else\Gin@nat@height\fi}
\makeatother
\setkeys{Gin}{width=\maxwidth,height=\maxheight,keepaspectratio}
\makeatletter
\def\fps@figure{htbp}
\makeatother
\usepackage{amsmath}
\usepackage{array}
\usepackage{natbib}
\usepackage{placeins}
\usepackage{orcidlink}
\usepackage{caption}
\usepackage{ulem}
\usepackage{enumitem}
\usepackage{rotating}

\ifLuaTeX
  \usepackage{selnolig}  
\fi
\IfFileExists{bookmark.sty}{\usepackage{bookmark}}{\usepackage{hyperref}}
\IfFileExists{xurl.sty}{\usepackage{xurl}}{} 
\hypersetup{
  pdftitle={Penalised regression with multiple sources of prior effects},
  hidelinks,
  pdfcreator={LaTeX via pandoc}}

\title{Penalised regression with multiple sources of prior effects}
\usepackage{etoolbox}
\makeatletter
\providecommand{\subtitle}[1]{
  \apptocmd{\@title}{\par {\large #1 \par}}{}{}
}
\makeatother
\subtitle{\(\qquad \qquad\) Armin Rauschenberger\(~^{1,*}\)
\orcidlink{0000-0001-6498-4801}, Zied Landoulsi\(~^{1}\)
\orcidlink{0000-0002-2327-3904}, \newline Mark A. van de
Wiel\(~^{2,3,\dagger}\) \orcidlink{0000-0003-4780-8472}, Enrico
Glaab\(~^{1,\dagger}\) \orcidlink{0000-0003-3977-7469}}
\author{}
\date{\vspace{-2.5em}16 December 2022}

\begin{document}
\maketitle

\(^1\)Luxembourg Centre for Systems Biomedicine (\textsc{lcsb}),
University of Luxembourg, Esch-sur-Alzette, Luxembourg. \(^2\)Department
of Epidemiology and Data Science (\textsc{eds}), Amsterdam University
Medical Centers (Amsterdam \textsc{umc}), Amsterdam, The Netherlands
\(^3\)Medical Research Council Biostatistics Unit (\textsc{mrc}
\textsc{bsu}), University of Cambridge, Cambridge, United Kingdom
\(^{*}\)To whom correspondence should be addressed. \(^{\dagger}\)Mark
A. van de Wiel and Enrico Glaab share senior authorship.

\textbf{In many high-dimensional prediction or classification tasks,
complementary data on the features are available, e.g.~prior biological
knowledge on (epi)genetic markers. Here we consider tasks with numerical
prior information that provide an insight into the importance (weight)
and the direction (sign) of the feature effects, e.g.~regression
coefficients from previous studies. We propose an approach for
integrating multiple sources of such prior information into penalised
regression. If suitable co-data are available, this improves the
predictive performance, as shown by simulation and application. The
proposed method is implemented in the R package `transreg'
(\url{https://github.com/lcsb-bds/transreg}).}

\textbf{keywords:} transfer learning; co-data; prior information; ridge
regression

\hypertarget{background}{%
\section{Background}\label{background}}

For many biomedical prediction or classification studies, there is a
previous study with a similar target and a similar high-dimensional
feature space, e.g.~hundreds of micro\textsc{rna}s (mi\textsc{rna}s),
thousands of genes, or millions of single-nucleotide polymorphisms
(\textsc{snp}s). Given a trained model from a previous study, we could
use it to obtain predicted values or predicted probabilities for the
study of interest, but these predictions are only reliable if the two
studies have the same target, the same features, and the same
population. However, we expect the feature-target effects from two
studies to be strongly correlated in more situations: slightly different
targets (e.g.~disease status vs disease stage), slightly different
features (e.g.~imperfectly overlapping feature space, different
measurement technique), slightly different populations
(e.g.~hospitalised vs non-hospitalised patients), or even different
modelling approaches (e.g.~simple regression vs multiple regression). As
it is challenging to estimate feature-target effects in high-dimensional
settings, it might be advantageous to use results from previous studies
as prior information for the study of interest.

Consider two prediction or classification problems, each one with a
target vector and a feature matrix (samples in the rows, features in the
columns). Suppose that both feature matrices cover the same features
(each column in the first matrix corresponds to a column in the second
matrix). In two special cases, the two problems reduce to a single
problem: (i) If both problems have the same target and concern samples
from the same population, they are in essence one
\text{`}single-target\text{'} problem (combine target vectors and
feature matrices by rows, respectively), potentially with batch effects.
(ii) If both problems concern the same samples, they are in essence one
\text{`}multi-target\text{'} problem (combine target vectors by columns,
feature matrices are the same). In other cases, however, the two
problems do not reduce to a single problem. Then we are in a potential
transfer learning setting (Table \ref{table_set}).

In such settings - two or more regression problems with related targets
and matched features - it might be possible to transfer information from
one problem to another. If the regression problems are sufficiently
related to each other, we expect their regression coefficients to be
correlated (positively or negatively). When fitting the regression model
of interest, we could therefore account for the estimated regression
coefficients from the other model. Transferring information on the
importance and the direction of the feature effects, we could
potentially increase the predictive performance.

\cite{Jiang2016} proposed the prior lasso to account for prior
information in high-dimensional predictive modelling. Their method
involves a preprocessing step and a weighting step. In the preprocessing
step, the prior information is used to predict the target from the
features. They present a solution for one set of prior effects from a
closely related study (multiplying the feature matrix by the prior
effects), but extensions to multiple sets of prior effects or loosely
related studies may be feasible. Let \(\boldsymbol{y}\) represent the
target and let \(\hat{\boldsymbol{y}}_{\text{prior}}\) represent the
fitted values based on the prior information. In the weighting step,
they minimise the penalised combined likelihood
\(L(\boldsymbol{x},\boldsymbol{y};\boldsymbol{\beta}) + \eta L(\boldsymbol{x},\hat{\boldsymbol{y}}_{\text{prior}};\boldsymbol{\beta}) - \rho(\lambda;\boldsymbol{\beta})\)
with respect to the coefficients \(\boldsymbol{\beta}\), where
\(\eta \geq 0\) (balance) and \(\lambda \geq 0\) (regularisation). If
the balancing hyperparameter \(\eta\) is larger than zero, the prior
predictions \(\hat{\boldsymbol{y}}_{\text{prior}}\) influence the
estimation of the parameters \(\boldsymbol{\beta}\).

\cite{Dhruba2021} proposed a transfer learning method based on
distribution mapping. Even if features or targets follow different
distributions in two data sets, it is possible to build a predictive
model using the first data set and make predictions for the second data
set. Requiring matched features and targets in the source data set and
unmatched features and targets in the target data set, their method
transfers (i) features from the target to the source domain and (ii)
predictions from the source to the target domain. By contrast, we
consider transfer learning settings with matched features and targets in
the target data set.

\cite{Tian2022} proposed and implemented transfer learning for ridge and
lasso regression. Their transfer learning algorithm involves two steps:
(i) Estimating common coefficients for the target data set and the
transferable source data sets (\(\hat{\boldsymbol{\omega}}\)). (ii)
Estimating the deviations from the common coefficients to the target
coefficients (\(\hat{\boldsymbol{\delta}}\)). Both steps together lead
to the estimated target coefficients
(\(\hat{\boldsymbol{\beta}} = \hat{\boldsymbol{\omega}} + \hat{\boldsymbol{\delta}}\)).
Before applying their transfer learning algorithm, \cite{Tian2022} apply
a transferable source detection algorithm to exclude source data sets
that are too different from the target data set. This avoids that
non-transferable sources render the common coefficients misleading for
the target data set (\text{`}negative transfer\text{'}). In the case of
lasso regularisation in the two steps, there is sparsity in the common
estimates as well as in the deviations from the common estimates to the
target estimates (and thereby also in the target estimates).

The method from \cite{Tian2022} requires not only the target data set
but also the source data set(s). However, data protection regulations or
restrictive data sharing policies might prevent researchers from
accessing a source data set, or the available storage or processing
capacity might be insufficient for analysing massive source data sets.
There is therefore a need for transfer learning methods that do not
require the source data but only the (anonymised) complementary data
(co-data) derived from the source data. Such methods allow us to exploit
summary statistics from external studies, e.g.~\(p\)-values and effect
sizes from a genome-wide association study (\textsc{gwas}), to increase
the predictive performance in the study of interest.

We propose a two-step transfer learning method, modelling with and
without co-data in the first step and combining different models in the
second step. Unless the source and target data sets are very similar,
the coefficients from the source data set(s) will not fit well to the
target data set. We therefore propose to calibrate these coefficients -
preserving their signs and their order - so that they can be transferred
from the source data set(s) to the target data set. Additionally, we
also estimate the coefficients directly from the target data set,
ignoring the co-data. The calibrated coefficients from the source data
set(s) as well as the estimated coefficients from the target data set
allow us to predict the outcome from the features. Finally, we combine
the linear predictors from the models with and without co-data and
calculate either predicted values (linear regression) or predicted
probabilities (logistic regression).

In a related transfer learning setting, prior information is only
available on the importance but not on the direction of the feature
effects, i.e.~with complementary data consisting of prior weights rather
than prior effects. In the generalised linear model framework, the
weighted lasso \citep{Bergersen2011}, the feature-weighted elastic net
\citep{Tay2022}, and penalised regression with differential shrinkage
\citep{Zeng2021} account for prior weights in the penalty function,
through feature-specific penalty factors or feature-specific
regularisation parameters. Adaptive group-regularised ridge regression
\citep{Wiel2016} is not only applicable to categorical co-data but also
to numerical co-data (prior weights), by the means of creating groups of
features from numerical co-data and forcing the group-penalties to be
monotonically decreasing. An extension from \cite{Nee2021} makes this
approach even more suitable for numerical co-data. For single sources of
co-data, it might be possible to extend these methods to prior
information on the importance as well as the direction of feature
effects by imposing sign constraints on the coefficients. Prior weights
have not only been exploited in regression analysis, e.g.~co-data
moderated random forests \citep{Beest2017} adapt the sampling
probabilities of the features to the prior weights.

\begin{table}
\caption{Abstract representation of the data set of interest (without asterisk, black) and an additional data set (with asterisk, grey). Single-target learning (left): same targets, same features, different samples (from one population). Multi-target learning (centre): different targets, same features, same samples. Transfer learning (right): same or different targets, matched features, different samples (from one or two populations).}
\label{table_set}
$$
\hspace{-1em}
\begin{array}{ccc}
\textit{Single-target learning} & \textit{Multi-target learning} & \textit{Transfer learning} \\
\begin{array}{ccc}
\left(
\begin{array}{c}
y_{1}^{\phantom{*}} \\
\vdots \\
y_{n}^{\phantom{*}} \\
\color{gray}{y_{1}^*} \\
\color{gray}{\vdots} \\
\color{gray}{y_{m}^*}
\end{array}
\right)
& \hspace{-1em}
\Leftarrow 
& \hspace{-1em}
\left(
\begin{array}{ccc}
x_{11}^{\phantom{*}} & \cdots & x_{1p}^{\phantom{*}} \\
\vdots & & \vdots \\
x_{n1}^{\phantom{*}} & \cdots & x_{np}^{\phantom{*}} \\
\color{gray}{x_{11}^*} & \color{gray}{\cdots} & \color{gray}{x_{1p}^*} \\
\color{gray}{\vdots} & & \color{gray}{\vdots} \\
\color{gray}{x_{m1}^*} & \color{gray}{\cdots} & \color{gray}{x_{mp}^*}
\end{array}
\right)
\end{array}
& \hspace{-1em}
\begin{array}{ccc}
\left(
\begin{array}{cc}
y_{1}^{\phantom{*}} & \color{gray}{y_{1}^*} \\
\vdots & \color{gray}{\vdots} \\
y_{n}^{\phantom{*}} & \color{gray}{y_{n}^*}
\end{array}
\right)
& \hspace{-1em}
\Leftarrow
& \hspace{-1em}
\left(
\begin{array}{ccc}
x_{11}^{\phantom{*}} & \cdots & x_{1p}^{\phantom{*}} \\
\vdots & & \vdots \\
x_{n1}^{\phantom{*}} & \cdots & x_{np}^{\phantom{*}} \\
\end{array}
\right)
\end{array}
& \hspace{-1em}
\begin{array}{ccc}
\left(
\begin{array}{c}
y_1^{\phantom{*}} \\
\vdots \\
y_{n}^{\phantom{*}}
\end{array}
\right)
& \hspace{-1em}
\Leftarrow
& \hspace{-1em}
\left(
\begin{array}{ccc}
x_{11}^{\phantom{*}} & \cdots & x_{1p}^{\phantom{*}} \\
\vdots & & \vdots \\
x_{n1}^{\phantom{*}} & \cdots & x_{np}^{\phantom{*}} \\
\end{array}
\right)
\\
& \hspace{-1em} & \hspace{-1em} ~~\Updownarrow~~ ~~\Updownarrow~~ ~~\Updownarrow~~
\\
\left(
\begin{array}{c}
\color{gray}{y_1^*} \\
\color{gray}{\vdots} \\
\color{gray}{y_m^*}
\end{array}
\right)
& \hspace{-1em}
\Leftarrow
& \hspace{-1em}
\left(
\begin{array}{ccc}
\color{gray}{x_{11}^*} & \color{gray}{\cdots} & \color{gray}{x_{1p}^*} \\
\color{gray}{\vdots} & & \color{gray}{\vdots} \\
\color{gray}{x_{m1}^*} & \color{gray}{\cdots} & \color{gray}{x_{mp}^*} \\
\end{array}
\right)
\end{array}
\end{array}
$$
\end{table}

\hypertarget{method}{%
\section{Method}\label{method}}

\hypertarget{model}{%
\subsection{Model}\label{model}}

Suppose one target and \(p\) features are available for \(n\) samples.
We index the samples by \(i\) in \(\{1,\ldots,n\}\) and the features by
\(j\) in \(\{1,\ldots,p\}\). Our aim is to estimate the generalised
linear model \[
\mathbb{E}[y_i] = h^{-1} \left( \beta_0 + \sum_{j=1}^{p} \beta_j x_{ij} \right)~.
\] For any sample \(i\), the model expresses the expected value of its
target (\(y_i\)) as a function of its features
(\(x_{i1},\ldots,x_{ip}\)). The link function \(h(\cdot)\) depends on
the family of distributions for the target (Gaussian: identity,
binomial: logit, Poisson: log). In the linear predictor, \(\beta_0\)
represents the unknown intercept, and \(\beta_j\) represents the unknown
slope of feature \(j\) (i.e.~the effect of the feature on the linear
predictor of the target). Given the estimated intercept
\(\hat{\beta}{}_0^\star\) and the estimated slopes
\(\{\hat{\beta}{}_1^\star,\ldots,\hat{\beta}{}_p^\star\}\), we could
predict the target of previously unseen samples: \[
\hat{y}_{i} = h^{-1} \left( \hat{\beta}{}_0^\star + \sum_{j=1}^{p} \hat{\beta}{}_j^\star x_{ij} \right)~.
\]

\hypertarget{co-data}{%
\subsection{Co-data}\label{co-data}}

Suppose \(m\) sources of co-data are available, indexed by \(k\) in
\(\{1,\ldots,m\}\). Let \(z_{jk}\) indicate the prior effect from source
\(k\) for feature \(j\). Our method is designed for quantitative co-data
that provide an insight into the importance (absolute value) and the
direction (sign) of the feature effects. Each set of prior effects
(\(-1 \leq \boldsymbol{z}_{\circ k} \leq 1\)) is assumed to be
positively correlated with the true coefficients
(\(\mathrm{cor}(\boldsymbol{z}_{\circ k},\boldsymbol{\beta}) > 0\)). For
any source of co-data, the prior effects may be re-scaled (not
re-centred), for example to the interval from \(-1\) to \(+1\). In other
words, the proposed method is invariant under multiplication of the
prior effects by a positive scalar
(\(\boldsymbol{z} \rightarrow c \times \boldsymbol{z}\) where
\(c > 0\)). We explain in the next section why this is important.

It might seem trivial to also allow for co-data that only provide an
insight into the importance but not the direction of the feature effects
(i.e.~prior weights instead of prior effects). Each set of prior weights
(\(0 \leq \boldsymbol{z}_{\circ k} \leq 1\)) is assumed to be positively
correlated with the true absolute coefficients
(\(\mathrm{cor}(\boldsymbol{z}_{\circ k},|\boldsymbol{\beta}|) > 0\)).
To obtain prior effects, one might want to assign the signs of the
Spearman correlation coefficients between the target and the features to
the prior weights. However, marginal effects and conditional effects can
have opposite signs. If we wanted to extend our approach to prior
weights, we would have to discover the signs inside the calibration
procedure (see below), which would be related to high-dimensional
regression with binary coefficients \citep{Gamarnik2017}.

\hypertarget{base-learners-with-co-data}{%
\subsection{Base-learners with
co-data}\label{base-learners-with-co-data}}

Suppose we are in a transfer learning setting with two prediction or
classification problems. For simplicity, we assume that the features do
not differ in scale between the two problems. For illustration, we
consider two artificial situations (where we would not use transfer
learning in practice): (i) If both problems concern the same target on
the same scale and the samples come from the same population, we could
use the estimated regression coefficients from one problem to make
predictions for the other problem. (ii) If the two problems concern the
same target on different scales, we could also recycle the estimated
regression coefficients, but we would have to adjust for the different
scales.

When transferring estimated regression coefficients from one problem to
another problem, it might not only be necessary to change their scale
but it might also be beneficial to change their shape. For example, it
might be that for one problem weak and strong effects matter, while for
the other problem only strong effects matter. We should therefore also
be able to make differences between small coefficients more or less
important than those between large coefficients. We propose two
calibration methods, namely exponential and isotonic calibration, to
adapt the prior information to the data. For each source of co-data
\(k\), both calibration methods estimate the model \[
\mathbb{E}[y_i] = h^{-1} \left( \alpha_k +  \sum_{j=1}^p \gamma_{jk} x_{ij} \right)~,
\] where the calibrated prior effects
\(\{\hat{\gamma}_{1k},\ldots,\hat{\gamma}_{pk}\}\) depend on the initial
prior effects \(\{z_{1k},\ldots,z_{pk}\}\). The difference between
exponential and isotonic scaling is how the former depend on the latter.

\begin{itemize}

\item exponential calibration: Let $\gamma_{jk}=\theta_k \mathrm{sign} (z_{jk}) |z_{jk}|{}^{\tau_k}$, for $j$ in $\{1,\ldots,p\}$, where the factor $\theta_k$ and the exponent $\tau_k$ are non-negative real numbers ($\theta_k \geq 0$ and $\tau_k \geq 0$). We first fit one simple non-negative regression for different values of $\tau_k$ (i.e. estimate $\alpha_k$ and $\theta_k$ given $\tau_k$), and then optimise $\tau_k$. Once $\theta_k$ and $\tau_k$ have been estimated, the initial prior effects $z_{jk}$ determine the final prior effects $\hat{\gamma}_{jk}=\hat{\theta}_k \mathrm{sign} (z_{jk}) |z_{jk}|^{\hat{\tau}_k}$, for all $j$ in $\{1,\ldots,p\}$. The estimated factor $\hat{\theta}_k$ and the estimated exponent $\hat{\tau}_k$ allow the model to change the scale and the shape of the prior effects. For example, $\hat{\theta}_k=0$ sets them to zero, $|\hat{\theta}_k|<1$ makes them smaller, $|\hat{\theta}_k|>1$ makes them larger, $\hat{\tau}_k=0$ sets them to the same value, $\hat{\tau}_k<1$ makes (absolutely) large ones more similar, and $\hat{\tau}_k>1$ makes (absolutely) small ones more similar. If one or more sets of prior effects might be negatively associated with the true coefficients, we could remove the non-negativity constraints from the simple regressions (allowing $\hat{\theta}_k<0$ to invert the signs of the prior effects).

\item isotonic calibration: We estimate $\{\gamma_{1k},\ldots,\gamma_{pk}\}$ under the constraint that the signs of the initial prior effects $z_{jk}$ determine the signs of the final prior effects $\hat{\gamma}_{jk}$ (i.e. $\hat{\gamma}_{jk}=0|z_{jk}=0$,
$\hat{\gamma}_{jk} \geq 0 | z_{jk} > 0$, $\hat{\gamma}_{jk} \leq 0 | z_{jk} < 0$) and under the constraint that the order of the initial prior effects determines the order of the final prior effects (i.e. $\hat{\gamma}_{jk} \geq \hat{\gamma}_{lk} | z_{jk} \geq z_{lk}$, $\hat{\gamma}_{jk} \leq \hat{\gamma}_{lk} | z_{jk} \leq z_{lk}$), for all $j$ and $l$ in $\{1,\ldots,p\}$. If one or more sets of prior effects might be negatively associated with the true coefficients, we could fit each model with these constraints and the inverted constraints, and then select the better fit.

To make optimisation more efficient, we rewrite the sign- and order-constrained problem as a sign-constrained problem (see Table~\ref{table_sub} in the Appendix). For each source of co-data, we order the columns of the feature matrix by increasing values of the prior effects. Suppose the first $q$ columns correspond to negative prior effects and the last $p-q$ columns correspond to non-negative prior effects. We take the cumulative sum of the feature columns from left to right for the former (columns $1$ to $q$) and from right to left for the latter (columns $p$ to $q+1$). We then estimate the coefficients on the left under non-positivity constraints, and those on the right under non-negativity constraints. Formally, the model equals
$$
\mathbb{E}[y_i]=h^{-1} \left( \alpha_k + \sum_{j=1}^{p} \delta_{jk} w_{ij} \right)~,
$$
where $w_{ij}=\sum_{l=1}^{j} x_{i(l)}$ and $\delta_{jk} \leq 0$ for $j$ in $\{1,\ldots,q\}$, and $w_{ij}=\sum_{l=j}^{p} x_{i(l)}$ and $\delta_{jk} \geq 0$ for $j$ in $\{q+1,\ldots,p\}$, with the subscript within brackets indicating the order of the prior effects. The linear predictor of the sign-constrained model, i.e. $\alpha_k + \sum_{j=1}^{{p}} \delta_{jk} w_{ij}$, is equivalent to the linear predictor of the order-constrained model, i.e. $\alpha_k + \sum_{j={1}}^p \gamma_{(j)k} x_{i(j)}$, because
$$
\begin{array}{ccccccc}
\sum\limits_{j=1}^q \delta_{jk} w_{ij}
&=& \sum\limits_{j=1}^q \delta_{jk} \left( \sum\limits_{l=1}^{j} x_{i(l)} \right)
&=& \sum\limits_{j=1}^q \left( \sum\limits_{l=j}^{q} \delta_{lk} \right) x_{i(j)}
&=& \sum\limits_{j=1}^q \gamma_{(j)k} x_{i(j)}~,
\\
\sum\limits_{j=q+1}^p \delta_{jk} w_{ij}
&=& \sum\limits_{j=q+1}^p \delta_{jk} \left( \sum\limits_{l=j}^{p} x_{i(l)} \right)
&=& \sum\limits_{j=q+1}^p \left( \sum\limits_{l=q+1}^{j} \delta_{lk} \right) x_{i(j)}
&=& \sum\limits_{j=q+1}^p \gamma_{(j)k} x_{i(j)}~.
\end{array}
$$
After estimating the coefficients of the sign-constrained model by maximum likelihood, we therefore estimate those of the order-constrained model by $\hat{\gamma}_{(j)k}=\sum_{l=j}^{q} \hat{\delta}_{lk}$ for $j$ in $\{1,\ldots,q\}$ and $\hat{\gamma}_{(j)k}=\sum_{l=q+1}^{j} \hat{\delta}_{lk}$ for $j$ in $\{q+1,\ldots,p\}$.

\end{itemize}

\begin{figure}
\centering
\includegraphics{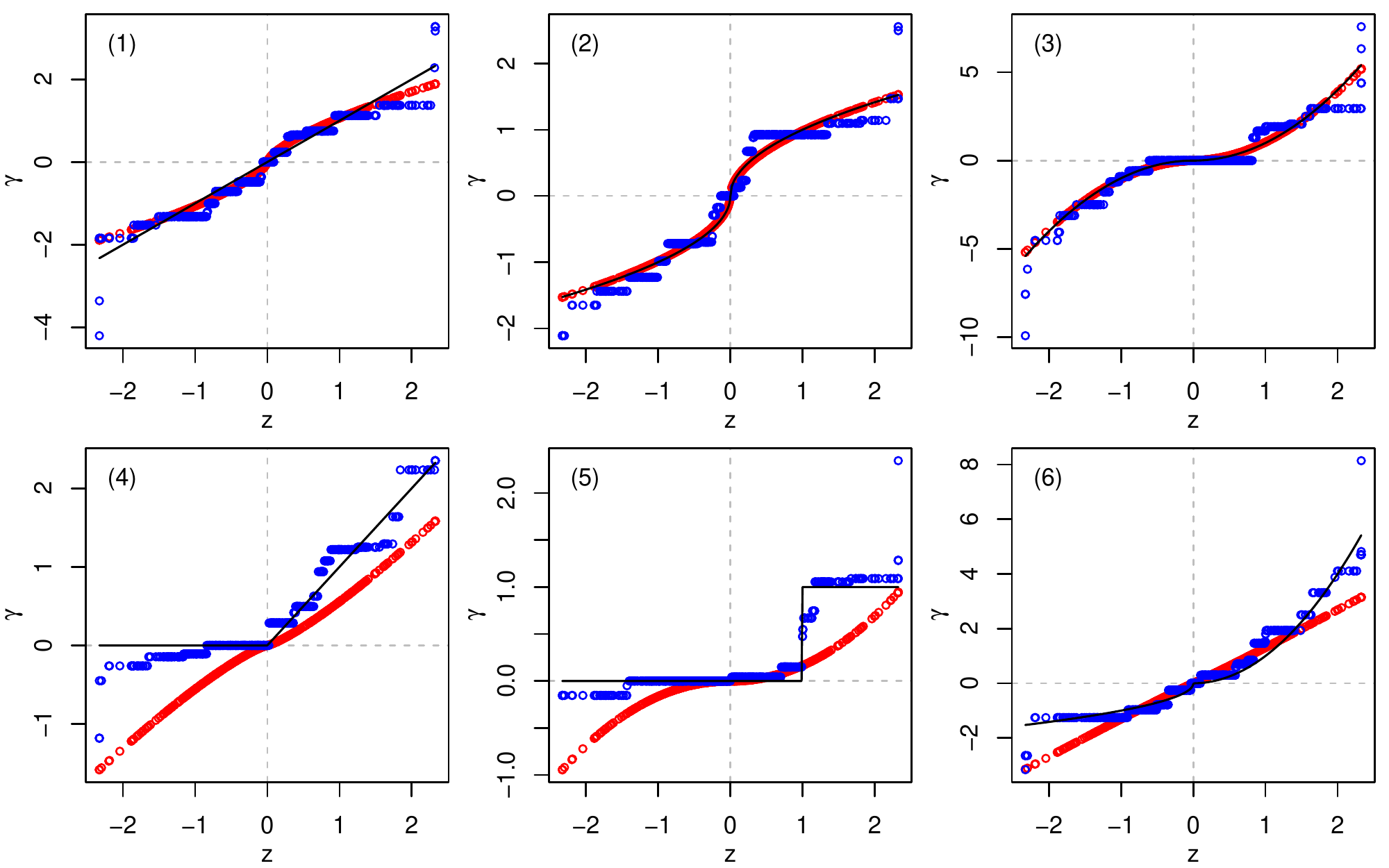}
\caption{Final prior effects $\boldsymbol{\gamma}$ ($y$-axis) against initial prior effects $\boldsymbol{z}$ ($x$-axis), under exponential calibration (red) and isotonic calibration (blue). The black line corresponds to perfect calibration ($\boldsymbol{\gamma}=\boldsymbol{\beta}$). We simulated the feature matrix $\boldsymbol{X}$ from a standard Gaussian distribution ($n=200$, $p=500$) and the initial prior effects $\boldsymbol{z}$ from a trimmed standard Gaussian distribution (trimmed below the $1\%$ and above the $99\%$ quantile). We set the true coefficients to (1)$~\boldsymbol{\beta}=\boldsymbol{z}$, (2)$~\boldsymbol{\beta}=\text{sign}(\boldsymbol{z}) \sqrt{|\boldsymbol{z}|}$, (3)$~\boldsymbol{\beta} = \text{sign}(\boldsymbol{z}) \boldsymbol{z}^2$, (4)$~\boldsymbol{\beta} = \mathbb{I}[\boldsymbol{z}>0] \boldsymbol{z}$, (5)$~\boldsymbol{\beta}=\mathbb{I}[\boldsymbol{z}>1]$, or (6)$~\boldsymbol{\beta} = - \mathbb{I}[\boldsymbol{z} \leq 0] \sqrt{|\boldsymbol{z}|} + \mathbb{I}[\boldsymbol{z}>0] \boldsymbol{z}^2$. And we simulated the response vector $\boldsymbol{y}$ from Gaussian distributions with the means $\boldsymbol{\eta}$ and the variance $\text{Var}(\boldsymbol{\eta})$, where  $\boldsymbol{\eta}=\boldsymbol{X}\boldsymbol{\beta}$. While exponential calibration performs slightly better in the first three scenarios (top), isotonic calibration performs much better in the last three scenarios (bottom).}
\label{fig_scale}
\end{figure}

While exponential calibration involves three unknown parameters, namely
the intercept \(\alpha_k\), the factor \(\theta_k\) and the exponent
\(\tau_k\), isotonic calibration involves \(1+p\) unknown parameters,
namely the intercept \(\alpha_k\) and the slopes
\(\boldsymbol{\gamma}_k=\{\gamma_{1k},\ldots,\gamma_{pk}\}\), for each
set of co-data. Figure \ref{fig_scale} shows the difference between
exponential and isotonic calibration in several empirically assessed
scenarios.

After calibration, we pre-assess the utility of each set of co-data. To
do this, we calculate the residuals (depending on the family of
distributions) between the fitted and the observed targets. We suggest
to retain a set of co-data only if the residuals are significantly
smaller than those of the intercept-only model (one-sided Wilcoxon
signed-rank test) at the nominal \(5\%\) level (\(p\)-value
\(\leq 0.05\)).

\FloatBarrier

\hypertarget{base-learners-without-co-data}{%
\subsection{Base-learners without
co-data}\label{base-learners-without-co-data}}

We also fit the model without any co-data. We estimate the coefficients
by maximising the penalised likelihood: \[
\hat{\boldsymbol{\beta}} = \underset{\boldsymbol{\beta}}{\mathrm{argmax}} \left\{ L(\boldsymbol{x};\boldsymbol{\beta}) - \rho(\lambda;\boldsymbol{\beta})\right\}~,
\] where \(L(\boldsymbol{x},\boldsymbol{\beta})\) is the likelihood and
\(\rho(\lambda;\boldsymbol{\beta})\) is the penalty. The likelihood
depends on the family of distributions (Gaussian, binomial, Poisson),
and the penalty can be the ridge (\(L_2\)) or the lasso (\(L_1\))
penalty. The penalty shrinks the squared (ridge) or absolute (lasso)
slopes \(\{\beta_1,\ldots,\beta_p\}\) towards zero (without penalising
the intercept \(\beta_0\)). We denote the estimated intercept by
\(\hat{\beta}_0\) and the estimated slopes by
\(\{\hat{\beta}_1,\ldots,\hat{\beta}_p\}\).

\hypertarget{cross-validation}{%
\subsection{Cross-validation}\label{cross-validation}}

We split the samples into ten folds to perform \(10\)-fold internal
cross-validation. In each iteration, we fit the models to nine included
folds and predict the target for the excluded fold.

Let the \(n \times m\) matrix \(\hat{\boldsymbol{H}}{}^{(0,\text{cv})}\)
represent the feature-dependent part of the cross-validated linear
predictors from the models with co-data. Specifically, the entry in row
\(i\) (sample) and column \(k\) (source of co-data) equals \[
\eta_{i{k}}^{(0,\text{cv})} = 0 \times \hat{\alpha}_k^{-\kappa(i)} + \sum_{j=1}^{p} \hat{\gamma}_{jk}^{-\kappa(i)} x_{ij}~,
\] where the superscript \(-\kappa(i)\) indicates that the (ignored)
intercept \(\alpha_k\) and the slopes \(\gamma_{jk}\) for \(j\) in
\(\{1,\ldots,p\}\) are estimated without using the fold of sample \(i\),
as in \cite{Rauschenberger2021b}.

The models without any co-data do not only have \(1+p\) unknown
parameters, namely the intercept \(\beta_0\) and the slopes
\(\{\beta_1,\ldots,\beta_p\}\), but also the unknown hyperparameter
\(\lambda\). In each iteration, we fit this model for a decreasing
sequence of \(100\) values for the regularisation parameter \(\lambda\),
indexed by \(l\) in \(\{1,\ldots,100\}\), using the computationally
efficient approach from
\citet[\href{https://CRAN.R-project.org/package=glmnet}{\texttt{glmnet}}]{Friedman2010}.

Accordingly, let the \(n \times 100\) matrix
\(\hat{\boldsymbol{H}}{}^{(1,\text{cv})}\) represent the cross-validated
linear predictors from the model without co-data. Specifically, the
entry in row \(i\) (sample) and column \(l\) (regularisation parameter)
equals \[
\eta_{il}^{(1,\text{cv})} = \hat{\beta}_0^{-\kappa(i),l} + \sum_{j=1}^p \hat{\beta}{}_j^{-\kappa(i),l} x_{ij}~,
\] where the superscripts \(-\kappa(i)\) and \(l\) indicate that the
intercept \(\beta_0\) and the slopes \(\{\beta_1,\ldots,\beta_p\}\) are
estimated without using the fold of sample \(i\) and given the
regularisation parameter \(\lambda_l\).

To optimise the predictive performance of the co-data independent model,
we would select the \(\lambda\) that minimises the cross-validated loss
(\(\lambda_{\text{min}}\)). As we base our predictions not only on the
co-data independent model but also on the co-data dependent model(s),
\(\lambda_{\text{min}}\) might be too small. The reason is that the
co-data might be informative to the extent that the co-data independent
model requires more penalisation. We could let the meta-learner select
the optimal \(\lambda\) from the whole sequence, but this might render
the inclusion and exclusion of co-data dependent models unstable. Our
ad-hoc solution is to include the optimal regularisation parameter for
the co-data independent model (\(\lambda_{\text{min}}\)) and a slightly
larger one (\(\lambda_{\text{1se}}\)). The latter is given by the
one-standard-error rule, which increases \(\lambda\) until the
cross-validated loss equals its minimum plus one standard error.

We concatenate \(\hat{\boldsymbol{H}}{}^{(0,\text{cv})}\) with the
columns of \(\hat{\boldsymbol{H}}{}^{(1,\text{cv})}\) that correspond to
\(\lambda_{\text{min}}\) and \(\lambda_{\text{1se}}\) to obtain the
\(n \times (m+2)\) matrix \(\hat{\boldsymbol{H}}{}^{(\text{cv})}\). The
first \(m\) columns correspond to the models with co-data, and the last
two columns correspond to the model without co-data.

\hypertarget{meta-learner}{%
\subsection{Meta-learner}\label{meta-learner}}

We combine the base-learners with and without co-data by stacked
generalisation \citep{Wolpert1992}, on the level of the linear
predictors \citep{Rauschenberger2021a}. In the meta-layer, we regress
the target on the cross-validated linear predictors from the base-layer:
\[
\mathbb{E}[y_i] = h^{-1} \left( \omega_0 + \sum_{k=1}^{m+2} \omega_k \hat{H}{}_{ik}^{(\text{cv})} \right)~.
\] Leaving the intercept unrestricted \((-\infty < \omega_0 < +\infty)\)
but imposing the lower bound zero on the slopes
\((\omega_1 \geq 0 , \ldots , \omega_{m+2} \geq 0)\), we estimate these
coefficients under lasso regularisation. Due to the feature selection
property of the lasso, a source of co-data can be excluded
(\(\hat{\omega}_k=0\)) or included (\(\hat{\omega}_k>0\)), where \(k\)
in \(\{1,\ldots,m\}\). Similarly, the models without co-data can be
excluded (\(\hat{\omega}_{k}=0\)) or included (\(\hat{\omega}_{k}>0\)),
where \(k=m+1\) for the model with \(\lambda_{\text{min}}\) and
\(k=m+2\) for the model with \(\lambda_{\text{1se}}\). The estimated
slopes function as weights for the co-data dependent models
(\(\hat{\omega}_1,\ldots,\hat{\omega}_m\)) and for the co-data
independent models (\(\hat{\omega}_{m+1},\hat{\omega}_{m+2}\)). Thus, we
do not only select sources but also weight them according to their
relevance.

\hypertarget{interpretation}{%
\subsection{Interpretation}\label{interpretation}}

The coefficients \(\hat{\boldsymbol{\beta}}_{\text{min}}\) and
\(\hat{\boldsymbol{\beta}}_{\text{1se}}\) give insight into the
feature-target effects estimated without co-data, with
\(\hat{\beta}_{\text{min},j}\) and \(\hat{\beta}_{\text{1se},j}\)
representing the effect of feature \(j\), where \(j\) in
\(\{1,\ldots,p\}\). The coefficients \(\hat{\boldsymbol{\omega}}\) give
insight into the importance of the sources of co-data, with
\(\hat{\omega}_k\) representing the importance of source \(k\), where
\(k\) in \(\{1,\ldots,m\}\). For a previously unseen sample \(i\), the
predicted value is: \[
\begin{aligned}
& \hat{y}_i = h^{-1} \left( \hat{\omega}_0 + \sum\limits_{k=1}^{m+2} \hat{\omega}_k \hat{H}_{ik} \right)
= h^{-1} \left( \hat{\beta}{}_0^{\star} + \sum\limits_{j=1}^p \hat{\beta}{}^{\star}_j x_{ij} \right)~,
\\
& \text{where} \quad
\hat{\beta}{}_0^\star = \hat{\omega}_0 + \hat{\omega}_{m+1} \hat{\beta}_{\text{min},0} + \hat{\omega}_{m+2} \hat{\beta}_{\text{1se},0}
\\
& \text{and} \quad
\hat{\beta}{}_j^\star = \left( \sum\limits_{k=1}^{m} \hat{\omega}_{k} \hat{\gamma}_{jk} \right) + \hat{\omega}_{m+1} \hat{\beta}_{\text{min},j} + \hat{\omega}_{m+2} \hat{\beta}_{\text{1se},j}~.
\end{aligned}
\] Thus, the estimated effect for a feature (\(\hat{\beta}{}_j^\star\))
is a weighted sum of estimated coefficients with co-data
(\(\hat{\gamma}_{j1},\ldots,\hat{\gamma}_{{j}m}\)) and the estimated
coefficients without co-data
(\(\hat{\beta}_{\text{min},j},\hat{\beta}_{\text{1se},j}\)).

Sparse models (few non-zero coefficients) are often considered to be
more interpretable than dense models (many non-zero coefficients). While
the original coefficients are dense
(\(\sum_{j=1}^{p} \mathbb{I}[\hat{\beta}{}_j \neq 0] = p\)) or sparse
(\(\sum_{j=1}^{p} \mathbb{I}[\hat{\beta}{}_j \neq 0] \ll p\)) depending
on the choice between ridge and lasso regularisation, the weights may
contain some zeros due to significance filtering or lasso regularisation
(\(\sum_{k=1}^{m+2} \mathbb{I}[\hat{\omega}_k \neq 0] \leq m+2\)). As
soon as one set of dense prior effects is selected, however, the
combined coefficients also become dense
(\(\sum_{j=1}^p \mathbb{I}[\hat{\beta}{}_j^\star \neq 0] \lesssim p\)).
This means that the feature selection property of the lasso is not
maintained. We should therefore choose between ridge and lasso
regularisation (i) to make the model without co-data more predictive or
interpretable (ii) or to make the model with co-data more predictive
(iii) but not to make the model with co-data more interpretable.

\FloatBarrier

\hypertarget{extension}{%
\subsection{Extension}\label{extension}}

In some applications, prior information might be reliable for some
features but unreliable for other features. Although the base-learners
with co-data might still be predictive, the meta-learner (weighted
average of the base-learners with and without co-data) might be not more
predictive than the base-learner without co-data. The reason is that the
meta-learner assigns the same weight to all prior effects, rather than
more weight to reliable prior effects and less weight to unreliable
prior effects. The same problem occurs if prior information is available
for some features but missing for other features. We therefore propose
an alternative approach for applications with partially informative
sources of co-data.

In the following, we use the term \text{`}meta-features\text{'} for the
cross-validated linear predictors from the base learners with co-data.
Each meta-feature - one column of the \(n \times m\) matrix
\(\hat{\boldsymbol{H}}{}^{(0,\text{cv})}\) - corresponds to one source
of co-data. In the meta-layer, we regress the target on the
meta-features and the base-features: \[
\mathbb{E}[y_i] = h^{-1} \left( \beta_0 + \sum_{k=1}^{m} \omega_k \hat{H}{}_{ik}^{(0,\text{cv})} + \sum_{j=1}^p \beta_j x_{ij} \right)~,
\] with non-negativity constraints for the weights for the meta-features
\((\omega_1 \geq 0, \ldots, \omega_m \geq 0)\) but without constraints
for the intercept (\(\beta_0\)) and the slopes for the base-features
\((\beta_1,\ldots,\beta_p)\).

We estimate the weights for the meta-features and the slopes for the
base-features using penalised maximum likelihood: \[
\{\hat{\boldsymbol{\omega}},\hat{\boldsymbol{\beta}}\} = \underset{\{\boldsymbol{\omega},\boldsymbol{\beta}\}}{\mathrm{argmax}} \left\{ L(\boldsymbol{x};\boldsymbol{\omega},\boldsymbol{\beta}) - \rho(\lambda;\boldsymbol{\beta})\right\}~,
\] where \(L(\boldsymbol{x};\boldsymbol{\omega},\boldsymbol{\beta})\) is
the likelihood and \(\rho(\lambda;\boldsymbol{\beta})\) is the penalty.
We do not penalise the weights for the meta-features (\(m \ll n\)) but
only the slopes for the base-features (\(p \gg n\)). The more sources of
co-data are available, the more it becomes necessary to penalise their
weights. But then the weights \(\boldsymbol{\omega}\) and the slopes
\(\boldsymbol{\beta}\) might need differential penalisation, for example
a lasso penalty for the meta-features (selection of sources) and a ridge
penalty for the base-features (many small effects). To make this
computationally efficient, we would need a fast cross-validation
procedure for multiple penalties \citep[cf.][]{Wiel2021} with
non-negativity constraints (meta-features) and mixed lasso and ridge
penalisation (meta-features vs base-features). This extension is
therefore only applicable in settings with few sources of co-data.

The predicted value for a previously unseen sample \(i\) is \[
\begin{aligned}
& \hat{y}_i = h^{-1} \left( \hat{\beta}_0 + \sum\limits_{k=1}^m \hat{\omega}_k \hat{H}{}_{ik}^{(0)} + \sum\limits_{j=1}^p \hat{\beta}_j x_{ij} \right) = h^{-1} \left( \hat{\beta}_0 + \sum\limits_{j=1}^p \hat{\beta}{}_j^\star x_{ij} \right)~,
\\
& \text{where} \quad \hat{\beta}{}_j^\star = \left( \sum\limits_{k=1}^{m} \hat{\omega}_{k} \hat{\gamma}_{jk} \right) + \hat{\beta}_j~.
\end{aligned}
\]

As the coefficients \(\boldsymbol{\beta}\) are shrunk towards zero but
the coefficients \(\boldsymbol{\omega}\) are not penalised, the combined
coefficients \(\boldsymbol{\beta}^\star\) are shrunk towards the
calibrated prior effects. When the regularisation parameter tends to
infinity (\(\lambda \to \infty\)), the estimated deviations from the
calibrated prior effects approach zero
\((\hat{\boldsymbol{\beta}} \to 0)\) and the combined estimates approach
the calibrated prior effects
\((\hat{\boldsymbol{\beta}}{}^{\star} \to \hat{\boldsymbol{\gamma}})\).
Lasso regularisation ensures sparsity in the deviations from the
calibrated prior effects
(\(\sum_{j=1}^p \mathbb{I} [\hat{\beta}_j \neq 0] \ll p\)) - in contrast
to ridge regularisation - but not in the combined coefficients
(\(\sum_{j=1}^p \mathbb{I} [\beta_j^\star \neq 0] \lesssim p\)). As the
combined coefficients may deviate more from unreliable than from
reliable calibrated prior effects, this extension is suitable for
partially informative co-data. As opposed to \text{`}standard
stacking\text{'}, we refer to this extension as \text{`}simultaneous
stacking\text{'}.

\FloatBarrier

\hypertarget{simulation}{%
\section{Simulation}\label{simulation}}

We performed two simulation studies to compare the predictive
performance between our transfer learning method and the one from
\cite{Tian2022}. In contrast to the method from \cite{Tian2022}, which
requires the feature-target effects in the target and the source data
set(s) to be \textit{positively correlated} and on the
\textit{same scale}
(i.e.~\(\boldsymbol{\beta}_{\text{target}} \approx \boldsymbol{\beta}_{\text{source}}\)),
our method also allows for negatively correlated effects and for effects
on different scales
(i.e.~\(\boldsymbol{\beta}_{\text{target}} \approx c \times \boldsymbol{\beta}_{\text{source}}\)).
Although it is possible to overcome this restriction by inverting the
target (Gaussian:
\(\boldsymbol{y}_{\text{source}} \to - \boldsymbol{y}_{\text{source}}\),
binomial:
\(\boldsymbol{y}_{\text{source}} \to 1-\boldsymbol{y}_{\text{source}}\)),
by re-scaling the target (Gaussian:
\(\boldsymbol{y}_{\text{source}} \to 1/c \times \boldsymbol{y}_{\text{source}}\)),
or by inverting or re-scaling the features
(\(\boldsymbol{X}_{\text{source}} \to c \times \boldsymbol{X}_{\text{source}}\)),
we believe it is more user-friendly to directly allow for negative
correlations and different scales. To ensure a fair comparison between
the two methods, we simulate positively correlated effects on the same
scale. Furthermore, although the method from \cite{Tian2022} is in
theory also suitable for mixed response types, the current version of
the related \texttt{R} package
\href{https://CRAN.R-project.org/package=glmtrans}{\texttt{glmtrans}}
requires the source data set(s) and the target data set to have the same
response type (Gaussian, binomial, Poisson). We therefore always
simulate the same response type in the source and target domains.

\hypertarget{external-simulation}{%
\subsection{External simulation}\label{external-simulation}}

We use the simulation approach from \cite{Tian2022}. In each iteration,
we call the function \texttt{glmtrans::models} with the arguments (1)
family of distributions: \texttt{family="gaussian"} (default) or
\texttt{family="binomial"}, (2) source or target data sets:
\texttt{type="all"} (default), (3) difference between source and target
coefficients: \texttt{h=5} (default) or \texttt{h=250}, (4) number of
source data sets: \texttt{K=5} (default), (5) sample size for target
data set: \texttt{n.target=100} (default), (6) sample size for each
source data set: \texttt{n.source=150} (default), (7) number of non-zero
coefficients: \texttt{s=15} (default) or \texttt{s=50}, (8) number of
features: \texttt{p=$1\,000$} (default), number of transferable source
data sets: \texttt{Ka=1}, \texttt{Ka=3} or \texttt{Ka=K=5} (default).

The simulation from \cite{Tian2022} involves the following steps:

\begin{itemize}

\item Features: The correlation between features $i$ and $j$ is set to $\Sigma_{ij}=0.5^{|i-j|}$, where $i$ and $j$ in $\{1,\ldots,p\}$. Let $\boldsymbol{\Sigma}$ represent the correlation matrix and let $\boldsymbol{\Sigma}=\boldsymbol{R}^\intercal \boldsymbol{R}$ represent its Cholesky decomposition, where $\boldsymbol{R}$ is an upper triangular matrix. For the target data set ($n_0=100$) and each source data set ($n_1=...=n_5=150$), the $n_0 \times p$ matrix $\boldsymbol{X}_0=\boldsymbol{E}_0 \boldsymbol{R}$ and the $n_k \times p$ matrices $\boldsymbol{X}_k=\boldsymbol{E}_k \boldsymbol{R}$ for $k$ in $\{1,\ldots,5\}$ represent the features, where the $n_0 \times p$ matrix $\boldsymbol{E}_{0}$ and the $n_k \times p$ matrices $\boldsymbol{E}_{k}$ contain Gaussian noise.

\item Coefficients: Let $\beta_j$ represent the effect of feature $j$, for $j$ in $\{1,\ldots,p\}$, and denote the $p$-dimensional coefficient vectors by $\boldsymbol{\beta}_0$ for the target data set and $\{\boldsymbol{\beta}_1,\ldots,\boldsymbol{\beta}_5\}$ for the source data sets. For the \textit{target} data set, the first $s$ elements are set to $\beta_j=0.5$ (causal) and the last $p-s$ elements are set to $\beta_j=0$ (non-causal). For \textit{transferable source} data sets, the first $s$ elements are set to $\beta_j = 0.5 + (-1)^{z_j} h/p$ and the last $p-s$ elements are set to $\beta_j= (-1)^{z_j} h/p$, where $z_j$ is a realisation of $z_j \sim \text{Bernoulli}(0.5)$. For \textit{non-transferable source} data sets, the first $s$ elements are set to $\beta_j = (-1)^{z_j} 2 h/p$ (non-causal), the next $s$ elements are set to $\beta_j = 0.5 + (-1)^{z_j} 2 h/p$ (causal), and the last $p - 2s$ elements are a random sample of $s$ causal and $p-3s$ non-causal elements generated in the same way. To obtain a non-transferable source, it would be sufficient to randomly select causal elements rather than inverting causal and non-causal elements (indices $1$ to $2s$).

\item Targets: In the \textit{Gaussian} case, the target vector is the $n$-dimensional vector $\boldsymbol{y}_0= \boldsymbol{X}_0 \boldsymbol{\beta}_0 + \boldsymbol{\epsilon}_0$ for the target data set, and the $n$-dimensional vector $\boldsymbol{y}{}_k=0.5 + \boldsymbol{X}{}_k\boldsymbol{\beta}{}_k + \boldsymbol{\epsilon}_k$ for the source data sets, where the $n$-dimensional vectors $\{\boldsymbol{\epsilon}_0,\ldots,\boldsymbol{\epsilon}_5\}$ contain Gaussian noise. In the \textit{binomial} case, let $\boldsymbol{p}_0=1/(1+\exp{(-\boldsymbol{X}_0 \boldsymbol{\beta}{}_0)})$ for the target data set and $\boldsymbol{p}{}_k=1/(1+\exp(-0.5-\boldsymbol{X}_k\boldsymbol{\beta}{}_k))$ for the source data sets. The $n$-dimensional vectors $\{\boldsymbol{y}_0,\ldots,\boldsymbol{y}_5\}$ are the target vectors, with each element following a Bernoulli distribution with the probability given by $\{\boldsymbol{p}_0,\ldots,\boldsymbol{p}_5\}$.

\end{itemize}

\hypertarget{internal-simulation}{%
\subsection{Internal simulation}\label{internal-simulation}}

The simulation from \cite{Tian2022} uses the same effect size for all
causal features and a decreasing correlation structure with a fixed
base. We therefore designed our own data-generating mechanism (i) to
simulate different effect sizes for different causal features
(\(\beta_j \in \mathbb{R}\) instead of \(\beta_j \in \{0,0.5\}\)) and
(ii) to vary the strength of correlation between features
(\(\sigma_{ij}=\rho_x{}^{|i-j|}\) instead of
\(\sigma_{ij}=0.5^{|i-j|}\)).

Our simulation involves the following steps:

\begin{itemize}

\item Features: Setting the mean of feature $i$ to $\mu_i=0$, the variance of feature $i$ to $\sigma_{ii}=1$, and the covariance between features $i$ and $j$ to $\sigma_{ij}=\rho_x^{|i-j|}$, for all $i$ and $j$ in $\{1,\ldots,p\}$, we simulate multiple feature matrices from the multivariate Gaussian distribution with mean vector $\boldsymbol{\mu}$ and covariance matrix $\boldsymbol{\Sigma}$, namely the $n_0 \times p$ feature matrix $\boldsymbol{X}_0$ for the target data set, and the $n_{1/2/3} \times p$ feature matrices $\{\boldsymbol{X}_1,\boldsymbol{X}_2,\boldsymbol{X}_3\}$ for the source data sets ($n_0=100, n_1=n_2=n_3=150, p=1\,000$).

\item Coefficients: Setting the mean and the variance for data set $k$ to $\mu_k=0$ and $\sigma_{kk}=1$, for all $k$ in $\{0,1,2,3\}$, and the covariance between data sets $k$ and $l$ to $\sigma_{kl}=0$ if either $k$ or $l$ equals $1$, or to $\sigma_{kl}=\rho_{\beta}$ if both $k$ and $l$ are in $\{0,2,3\}$, we simulate two $p \times 4$ matrices from the multivariate Gaussian distribution with mean vector $\boldsymbol{\mu}$ and covariance matrix $\boldsymbol{\Sigma}$, namely $\boldsymbol{B}_1$ and $\boldsymbol{B}_2$. We define the coefficients as $\boldsymbol{B}=\boldsymbol{B}_1 \mathbb{I}[\boldsymbol{B}_2 > \phi^{-1}(1-\pi)]$, where $\phi$ is the Gaussian cumulative distribution function and $\pi$ equals $0.2$ (dense) or $0.05$ (sparse), and denote the $p$-dimensional coefficient vectors by $\boldsymbol{\beta}_0$ for the target data set and by $\{\boldsymbol{\beta}_1,\boldsymbol{\beta}_2,\boldsymbol{\beta}_3\}$ for the source data sets. While one set of coefficients is non-transferable ($\boldsymbol{\beta}_1$), we transform the transferable sets of coefficients with $\boldsymbol{\beta}_2 \to \mathrm{sign}(\boldsymbol{\beta}_2) |\boldsymbol{\beta}_2|^2$ and $\boldsymbol{\beta}_3 \to \mathrm{sign}(\boldsymbol{\beta}_3) \sqrt{|\boldsymbol{\beta}_3|}$.

\item Targets: For the target data set, we compute $\boldsymbol{z}_0=\boldsymbol{X}_0 \boldsymbol{\beta}_0$ and standardise $\boldsymbol{z}_0$ to obtain $\boldsymbol{z}_0^*$. For the source data sets, we proceed similarly to obtain $\{\boldsymbol{z}_1^*,\boldsymbol{z}_2^*,\boldsymbol{z}_3^*\}$. The simulated targets equal $\boldsymbol{y}_k = h^{-1}(\sqrt{w} \boldsymbol{z}_k^* + \sqrt{1-w} \boldsymbol{\epsilon}_k)$, where $h(\cdot)$ is a link function and $\boldsymbol{\epsilon}_k$ follows a standard Gaussian distribution, for $k$ in $\{0,\ldots,3\}$. Given $0 \leq w \leq 0$, we have $\mathrm{Var}(\sqrt{w} \boldsymbol{z}_k^* + \sqrt{1-w} \boldsymbol{\epsilon}_k)=w \mathrm{Var}(\boldsymbol{z}_k^*) + (1-w) Var(\boldsymbol{\epsilon}_k)=1$. While $h(\cdot)$ is the identity link in the Gaussian case, it is the logit link in the binomial case, where the simulated probabilities are rounded to simulated classes.

\end{itemize}

\hypertarget{simulation-results}{%
\subsection{Simulation results}\label{simulation-results}}

In addition to the target data set, the method from \cite{Tian2022}
requires the source data sets, while our method requires the prior
effects derived from the source data sets. Since the two methods have
different requirements, we first simulate the source data sets (for the
competing method) and then derive the prior effects from the simulated
source data sets (for the proposed method). As prior effects for the
proposed method, we use the estimated regression coefficients from the
source data sets. We choose the type of regularisation for both methods
subject to the simulation setting, namely ridge regularisation for dense
settings
(\href{https://CRAN.R-project.org/package=glmtrans}{\texttt{glmtrans}}:
\(\alpha=0\); \texttt{transreg}:
\(\alpha_{\text{source}}=\alpha_{\text{target}}=0\)) and lasso or
lasso-like elastic net regularisation for sparse settings
(\href{https://CRAN.R-project.org/package=glmtrans}{\texttt{glmtrans}}:
\(\alpha=1\); \texttt{transreg}: \(\alpha_{\text{source}}=0.95\),
\(\alpha_{\text{target}}=1\)). The idea of the lasso-like elastic net
regularisation is to render the prior information more stable.

In each simulation setting, we simulate \(100\) training samples and
\(10\,000\) testing samples (hold-out) for the target data set. Tables
\ref{sim_ext} and \ref{sim_int} show the testing loss in the external
and internal simulation, under exponential and isotonic calibration. We
observe that transfer learning with
\href{https://CRAN.R-project.org/package=glmtrans}{\texttt{glmtrans}}
and \texttt{transreg} leads to an improvement with respect to
\href{https://CRAN.R-project.org/package=glmnet}{\texttt{glmnet}}.
Comparing the two different calibration approaches and the two different
stacking approaches, we do not observe systematic differences.

\begin{table}[ht]
\centering
\caption{Testing loss in external simulation, as a percentage of the one from prediction by the mean. Settings: number of transferable source data sets ($K_a$), differences between source and target coefficients ($h$), dense setting with ridge regularisation ($s=50$, $\alpha=0$) or sparse setting with lasso regularisation ($s=15$, $\alpha=1$), family of distribution (`gaussian' or `binomial'). These parameters determine (i) the average Pearson correlations among the features in the target data set ($\bar{\rho}_x$) and (ii) the maximum Pearson correlation between the coefficients in the target data set and the coefficients in the source data sets ($\max(\hat{\rho}_{\beta})$). Methods: regularised regression (\texttt{glmnet}), competing transfer learning method (\texttt{glmtrans}), proposed transfer learning method (\texttt{transreg}) with exponential/isotonic calibration and standard/simultaneous stacking. In each setting, the colour black highlights methods that are more predictive than regularised regression without transfer learning (\texttt{glmnet}), and an underline highlights the most predictive method. \label{sim_ext}} 
\begin{tabular}{|rrrrrr|rrrrrr|}
  \hline
$K_a$ & $h$ & $\alpha$ & family & $\bar{\rho}_{x}$ & $\max(\hat{\rho}_{\beta})$ & \texttt{glmnet} & \texttt{glmtrans} & \texttt{exp.sta} & \texttt{exp.sim} & \texttt{iso.sta} & \texttt{iso.sim} \\ 
  \hline
  1 &   5 &   0 & gaussian & 0.01 & 1.00 & ~\textcolor{gray}{73.6} & ~50.3 & ~37.3 & ~35.7 & ~31.6 & ~\underline{29.7} \\ 
    3 &   5 &   0 & gaussian & 0.01 & 1.00 & ~\textcolor{gray}{"} & ~38.9 & ~20.0 & ~18.2 & ~15.1 & ~\underline{13.6} \\ 
    5 &   5 &   0 & gaussian & 0.01 & 1.00 & ~\textcolor{gray}{"} & ~23.7 & ~13.8 & ~12.7 & ~10.3 & ~\underline{ 9.5} \\ 
    1 & 250 &   0 & gaussian & 0.01 & 0.40 & ~\textcolor{gray}{"} & ~\underline{39.5} & ~61.7 & ~65.2 & ~59.7 & ~58.4 \\ 
    3 & 250 &   0 & gaussian & 0.01 & 0.40 & ~\textcolor{gray}{"} & ~\underline{36.0} & ~48.0 & ~46.0 & ~42.6 & ~42.0 \\ 
    5 & 250 &   0 & gaussian & 0.01 & 0.40 & ~\textcolor{gray}{"} & ~34.1 & ~38.4 & ~36.7 & ~30.4 & ~\underline{27.8} \\ 
    1 &   5 &   1 & gaussian & 0.01 & 1.00 & ~\textcolor{gray}{23.3} & ~\underline{12.7} & ~14.3 & ~16.3 & ~13.8 & ~14.5 \\ 
    3 &   5 &   1 & gaussian & 0.01 & 1.00 & ~\textcolor{gray}{"} & ~\underline{ 9.7} & ~11.0 & ~11.6 & ~10.2 & ~10.0 \\ 
    5 &   5 &   1 & gaussian & 0.01 & 1.00 & ~\textcolor{gray}{"} & ~\underline{ 9.6} & ~10.8 & ~11.4 & ~10.0 & ~ 9.8 \\ 
    1 & 250 &   1 & gaussian & 0.01 & 0.26 & ~\textcolor{gray}{"} & ~\textcolor{gray}{58.4} & ~19.3 & ~\textcolor{gray}{28.4} & ~\underline{18.8} & ~\textcolor{gray}{28.7} \\ 
    3 & 250 &   1 & gaussian & 0.01 & 0.28 & ~\textcolor{gray}{"} & ~\textcolor{gray}{34.5} & ~\underline{18.7} & ~\textcolor{gray}{28.6} & ~18.8 & ~\textcolor{gray}{32.1} \\ 
    5 & 250 &   1 & gaussian & 0.01 & 0.28 & ~\textcolor{gray}{"} & ~\textcolor{gray}{34.5} & ~\underline{18.8} & ~\textcolor{gray}{29.3} & ~19.2 & ~\textcolor{gray}{35.9} \\ 
    1 &   5 &   0 & binomial & 0.01 & 1.00 & ~\textcolor{gray}{93.0} & ~91.7 & ~79.4 & ~81.3 & ~\underline{76.0} & ~76.0 \\ 
    3 &   5 &   0 & binomial & 0.01 & 1.00 & ~\textcolor{gray}{"} & ~85.2 & ~\underline{63.7} & ~65.9 & ~64.7 & ~65.0 \\ 
    5 &   5 &   0 & binomial & 0.01 & 1.00 & ~\textcolor{gray}{"} & ~79.4 & ~62.2 & ~64.8 & ~\underline{61.9} & ~63.1 \\ 
    1 & 250 &   0 & binomial & 0.01 & 0.39 & ~\textcolor{gray}{"} & ~89.5 & ~90.6 & ~89.7 & ~87.4 & ~\underline{85.4} \\ 
    3 & 250 &   0 & binomial & 0.01 & 0.43 & ~\textcolor{gray}{"} & ~88.6 & ~80.5 & ~82.3 & ~\underline{75.8} & ~76.4 \\ 
    5 & 250 &   0 & binomial & 0.01 & 0.44 & ~\textcolor{gray}{"} & ~84.8 & ~82.5 & ~84.7 & ~77.8 & ~\underline{77.5} \\ 
    1 &   5 &   1 & binomial & 0.01 & 1.00 & ~\textcolor{gray}{77.4} & ~73.4 & ~65.6 & ~70.9 & ~\underline{65.1} & ~70.5 \\ 
    3 &   5 &   1 & binomial & 0.01 & 1.00 & ~\textcolor{gray}{"} & ~63.9 & ~60.5 & ~60.8 & ~\underline{58.3} & ~61.0 \\ 
    5 &   5 &   1 & binomial & 0.01 & 1.00 & ~\textcolor{gray}{"} & ~60.2 & ~57.3 & ~58.4 & ~57.6 & ~\underline{55.8} \\ 
    1 & 250 &   1 & binomial & 0.01 & 0.20 & ~\textcolor{gray}{"} & ~\textcolor{gray}{85.5} & ~\underline{77.1} & ~\textcolor{gray}{77.4} & ~77.1 & ~\textcolor{gray}{77.4} \\ 
    3 & 250 &   1 & binomial & 0.01 & 0.25 & ~\textcolor{gray}{"} & ~\textcolor{gray}{79.6} & ~\underline{77.1} & ~\textcolor{gray}{77.4} & ~77.1 & ~\textcolor{gray}{77.4} \\ 
    5 & 250 &   1 & binomial & 0.01 & 0.26 & ~\textcolor{gray}{"} & ~76.7 & ~77.1 & ~\textcolor{gray}{77.4} & ~\underline{76.5} & ~\textcolor{gray}{82.3} \\ 
   \hline
\end{tabular}
\end{table}

\begin{table}[ht]
\centering
\caption{Testing loss in internal simulation, as a percentage of the one from prediction by the mean. Settings: correlation parameter for features ($\rho_x$), correlation parameter for coefficients ($\rho_{\beta}$), dense setting with ridge regularisation ($\pi=20\%$, $\alpha=0$) or sparse setting with lasso regularisation ($\pi=5\%$, $\alpha=1$), family of distribution (`gaussian' or `binomial'). These parameters determine (i) the average Pearson correlations among the features in the target data set ($\bar{\rho}_x$) and (ii) the maximum Pearson correlation between the coefficients in the target data set and the coefficients in the source data sets ($\max(\hat{\rho}_{\beta})$). Methods: regularised regression (\texttt{glmnet}), competing transfer learning method (\texttt{glmtrans}), proposed transfer learning method (\texttt{transreg}) with exponential/isotonic calibration and standard/simultaneous stacking. In each setting, the colour black highlights methods that are more predictive than regularised regression without transfer learning (\texttt{glmnet}), and an underline highlights the most predictive method. \label{sim_int}} 
\begin{tabular}{|rrrrrr|rrrrrr|}
  \hline
$\rho_x$ & $\rho_\beta$ & $\alpha$ & family & $\bar{\rho}_{x}$ & $\max(\hat{\rho}_{\beta})$ & \texttt{glmnet} & \texttt{glmtrans} & \texttt{exp.sta} & \texttt{exp.sim} & \texttt{iso.sta} & \texttt{iso.sim} \\ 
  \hline
0.95 & 0.70 &   0 & gaussian & 0.04 & 0.44 & ~\underline{\textcolor{gray}{ 76.5}} & ~\textcolor{gray}{ 79.7} & ~\textcolor{gray}{82.5} & ~\textcolor{gray}{79.5} & ~\textcolor{gray}{82.5} & ~\textcolor{gray}{79.5} \\ 
  0.99 & 0.70 &   0 & gaussian & 0.18 & 0.43 & ~\underline{\textcolor{gray}{ 63.8}} & ~\textcolor{gray}{ 68.3} & ~\textcolor{gray}{67.2} & ~\textcolor{gray}{65.7} & ~\textcolor{gray}{65.1} & ~\textcolor{gray}{64.4} \\ 
  0.95 & 0.85 &   0 & gaussian & 0.04 & 0.62 & ~\textcolor{gray}{ 78.0} & ~\textcolor{gray}{ 87.9} & ~75.6 & ~75.6 & ~75.0 & ~\underline{74.5} \\ 
  0.99 & 0.85 &   0 & gaussian & 0.18 & 0.64 & ~\textcolor{gray}{ 63.7} & ~\textcolor{gray}{ 64.2} & ~63.1 & ~\underline{63.0} & ~63.0 & ~63.2 \\ 
  0.95 & 0.99 &   0 & gaussian & 0.04 & 0.87 & ~\textcolor{gray}{ 70.3} & ~\textcolor{gray}{ 74.1} & ~65.9 & ~\underline{65.5} & ~65.6 & ~66.2 \\ 
  0.99 & 0.99 &   0 & gaussian & 0.18 & 0.87 & ~\textcolor{gray}{ 62.7} & ~\textcolor{gray}{ 64.4} & ~58.4 & ~59.7 & ~\underline{58.1} & ~61.1 \\ 
  0.95 & 0.70 &   1 & gaussian & 0.04 & 0.28 & ~\textcolor{gray}{ 78.6} & ~\underline{ 78.3} & ~\textcolor{gray}{78.6} & ~78.4 & ~\textcolor{gray}{78.6} & ~78.4 \\ 
  0.99 & 0.70 &   1 & gaussian & 0.18 & 0.14 & ~\textcolor{gray}{ 59.4} & ~ 59.3 & ~59.0 & ~\textcolor{gray}{59.4} & ~\underline{58.5} & ~59.3 \\ 
  0.95 & 0.85 &   1 & gaussian & 0.04 & 0.39 & ~\textcolor{gray}{ 73.4} & ~\underline{ 70.7} & ~71.8 & ~72.3 & ~72.5 & ~72.9 \\ 
  0.99 & 0.85 &   1 & gaussian & 0.18 & 0.58 & ~\textcolor{gray}{ 65.0} & ~\textcolor{gray}{ 65.9} & ~\textcolor{gray}{66.2} & ~\textcolor{gray}{65.6} & ~64.2 & ~\underline{62.4} \\ 
  0.95 & 0.99 &   1 & gaussian & 0.04 & 0.87 & ~\textcolor{gray}{ 88.2} & ~ 78.0 & ~74.8 & ~78.4 & ~\underline{73.8} & ~76.2 \\ 
  0.99 & 0.99 &   1 & gaussian & 0.18 & 0.77 & ~\textcolor{gray}{ 65.0} & ~ 62.8 & ~61.5 & ~60.4 & ~60.2 & ~\underline{59.8} \\ 
  0.95 & 0.70 &   0 & binomial & 0.04 & 0.36 & ~\textcolor{gray}{ 90.2} & ~\textcolor{gray}{ 98.4} & ~\textcolor{gray}{90.9} & ~\textcolor{gray}{93.0} & ~\underline{90.0} & ~\textcolor{gray}{91.4} \\ 
  0.99 & 0.70 &   0 & binomial & 0.18 & 0.39 & ~\textcolor{gray}{ 84.9} & ~\textcolor{gray}{ 90.0} & ~\textcolor{gray}{85.7} & ~\underline{84.2} & ~\textcolor{gray}{86.1} & ~84.7 \\ 
  0.95 & 0.85 &   0 & binomial & 0.04 & 0.61 & ~\textcolor{gray}{ 90.4} & ~\underline{ 89.3} & ~90.2 & ~\textcolor{gray}{90.4} & ~90.3 & ~90.2 \\ 
  0.99 & 0.85 &   0 & binomial & 0.18 & 0.54 & ~\textcolor{gray}{ 83.3} & ~\underline{ 82.7} & ~\textcolor{gray}{83.8} & ~83.0 & ~\textcolor{gray}{84.0} & ~\textcolor{gray}{83.8} \\ 
  0.95 & 0.99 &   0 & binomial & 0.04 & 0.88 & ~\textcolor{gray}{ 94.9} & ~ 92.7 & ~91.5 & ~93.0 & ~\underline{90.7} & ~92.5 \\ 
  0.99 & 0.99 &   0 & binomial & 0.18 & 0.89 & ~\textcolor{gray}{ 78.4} & ~ 78.2 & ~77.3 & ~77.8 & ~\underline{77.2} & ~\textcolor{gray}{78.4} \\ 
  0.95 & 0.70 &   1 & binomial & 0.04 & 0.54 & ~\textcolor{gray}{ 94.0} & ~ 90.4 & ~89.8 & ~\underline{89.4} & ~90.2 & ~89.6 \\ 
  0.99 & 0.70 &   1 & binomial & 0.18 & 0.29 & ~\textcolor{gray}{ 92.0} & ~\textcolor{gray}{ 92.5} & ~\textcolor{gray}{94.3} & ~\textcolor{gray}{92.0} & ~\textcolor{gray}{92.1} & ~\underline{90.3} \\ 
  0.95 & 0.85 &   1 & binomial & 0.04 & 0.28 & ~\textcolor{gray}{105.5} & ~104.4 & ~\underline{96.6} & ~97.6 & ~96.6 & ~97.6 \\ 
  0.99 & 0.85 &   1 & binomial & 0.18 & 0.57 & ~\textcolor{gray}{ 86.8} & ~\underline{ 85.1} & ~85.9 & ~\textcolor{gray}{87.5} & ~86.2 & ~85.5 \\ 
  0.95 & 0.99 &   1 & binomial & 0.04 & 0.92 & ~\textcolor{gray}{100.0} & ~ 94.4 & ~90.1 & ~88.2 & ~87.3 & ~\underline{86.8} \\ 
  0.99 & 0.99 &   1 & binomial & 0.18 & 0.82 & ~\textcolor{gray}{ 89.9} & ~ 89.7 & ~89.4 & ~86.8 & ~86.5 & ~\underline{86.1} \\ 
   \hline
\end{tabular}
\end{table}

\FloatBarrier

\hypertarget{applications}{%
\section{Applications}\label{applications}}

\hypertarget{external-applications}{%
\subsection{External applications}\label{external-applications}}

First, we consider an adapted version of the application on cervical
cancer from \cite{Wiel2016}. The aim is to transfer information from a
methylation study with biopsy samples to another methylation study with
self-collected samples in order to better discriminate between low-grade
and high-grade precursor lesions. Specifically, we transfer the signs of
the effect sizes and the \(p\)-values from the source data set to the
target data set (\(n=44\) samples, \(p=9\,491\) features). We then
examine whether transfer learning increases the predictive performance
of ridge regression, which is more predictive than lasso regression in
this application. For comparison, we consider the methods from
\citet[\texttt{fwelnet}]{Tay2022} and
\citet[\href{https://CRAN.R-project.org/package=ecpc}{\texttt{ecpc}}]{Nee2021}.
While the proposed method exploits information on the importance and
direction of the effects (co-data:
\(-\mathrm{sign}(\mathrm{coef}) \log_{10} (p\text{-value})\)), the other
two methods only exploit information on their importance (co-data:
\(-\log_{10} (p\text{-value})\)). After \(10\) repetitions of
\(10\)-fold cross-validation, we observe that transfer learning (not
with exponential but with isotonic calibration) often increases the
predictive performance of ridge regression (\texttt{transreg.exp.sta}:
\(0/10\), \texttt{transreg.exp.sim}: \(4/10\),
\texttt{transreg.iso.sta}: \(7/10\), \texttt{transreg.iso.sim}:
\(10/10\), \texttt{fwelnet}: \(7/10\),
\href{https://CRAN.R-project.org/package=ecpc}{\texttt{ecpc}}:
\(5/10\)). We also observe that exploiting information on the importance
as well as the direction of the effects (\texttt{transreg}) is more
beneficial than exploiting information on the importance of the effects
only (\texttt{fwelnet},
\href{https://CRAN.R-project.org/package=ecpc}{\texttt{ecpc}}), as can
be seen in the mean change in cross-validated loss
(\texttt{transreg.exp.sta}: \(+6.56\%\), \texttt{transreg.exp.sim}:
\(+0.43\%\), \texttt{transreg.iso.sta}: \(-2.50\%\),
\texttt{transreg.iso.sim}: \(-9.25\%\), \texttt{fwelnet}: \(-0.12\%\),
\href{https://CRAN.R-project.org/package=ecpc}{\texttt{ecpc}}:
\(-1.52\%\)). Here, isotonic calibration outperforms exponential
calibration, and simultaneous stacking outperforms standard stacking. A
potential explanation for the large difference in performance between
exponential and isotonic calibration is that positive effects might be
more important than negative effects in this application, for a
biological reason (methylation increases the probability of cancer) and
a statistical reason (effects of overexpression are easier to detect
than those of underexpression). While exponential calibration behaves
symmetrically for negative and positive prior effects, isotonic
calibration can shrink negative prior effects towards zero.

Second, we consider an adapted version of the application on
pre-eclampsia from \cite{Tay2022}. Measurements of \(p=1\,125\) plasma
proteins are available for \(n=166\) patients at multiple time points
(\(48 \times 2 + 8 \times 3 + 20 \times 4 + 74 \times 5 + 16 \times 6 = 666\)).
The aim is to transfer information from late time points (gestational
age \(>20\) weeks) to early time points (gestational age \(\leq 20\)
weeks). We repeatedly split the patients into one source data set and
one target data set. Patients with only late time points are always in
the source data set, and other patients are randomly allocated to the
source and the target data set. (Note that this application is somewhat
artificial, as it might be better to drop transfer learning in favour of
using all earliest time points in the regression of interest.) Using the
source data set, we estimate two logistic regression models under ridge
regularisation, once using the early time points and once using all time
points. For each patient, all time points are assigned to the same
cross-validation fold, and the weight is split evenly among the time
points. We then use the two sets of estimated regression coefficients as
co-data for the target data set. In the regression for the target data
set, we only include the earliest time point of each patient. Using
\(10\)-fold cross-validation, we estimate the predictive performance of
ridge regression with and without transfer learning. After repeating
source-target splitting and cross-validation \(10\) times, we observe
that transfer learning tends to decrease the cross-validated logistic
deviance (\texttt{transreg.exp.sta}: \(8/10\),
\texttt{transreg.exp.sim}: \(7/10\), \texttt{transreg.iso.sta}:
\(7/10\), \texttt{transreg.iso.sim}: \(9/10\), \texttt{fwelnet}:
\(5/10\), \href{https://CRAN.R-project.org/package=ecpc}{\texttt{ecpc}}:
\(6/10\)). It is more beneficial to share information not only on the
importance but also the direction of the effects, according to the mean
change in cross-validated logistic deviance (\texttt{transreg.exp.sta}:
\(-2.61\%\), \texttt{transreg.exp.sim}: \(-4.29\%\),
\texttt{transreg.iso.sta}: \(-3.67\%\), \texttt{transreg.iso.sim}:
\(-8.33\%\), \texttt{fwelnet}: \(+0.04\%\),
\href{https://CRAN.R-project.org/package=ecpc}{\texttt{ecpc}}:
\(-6.87\%\)). Simultaneous stacking again outperforms standard stacking,
but exponential and isotonic calibration show a similar performance.

\begin{figure}
\centering
\includegraphics{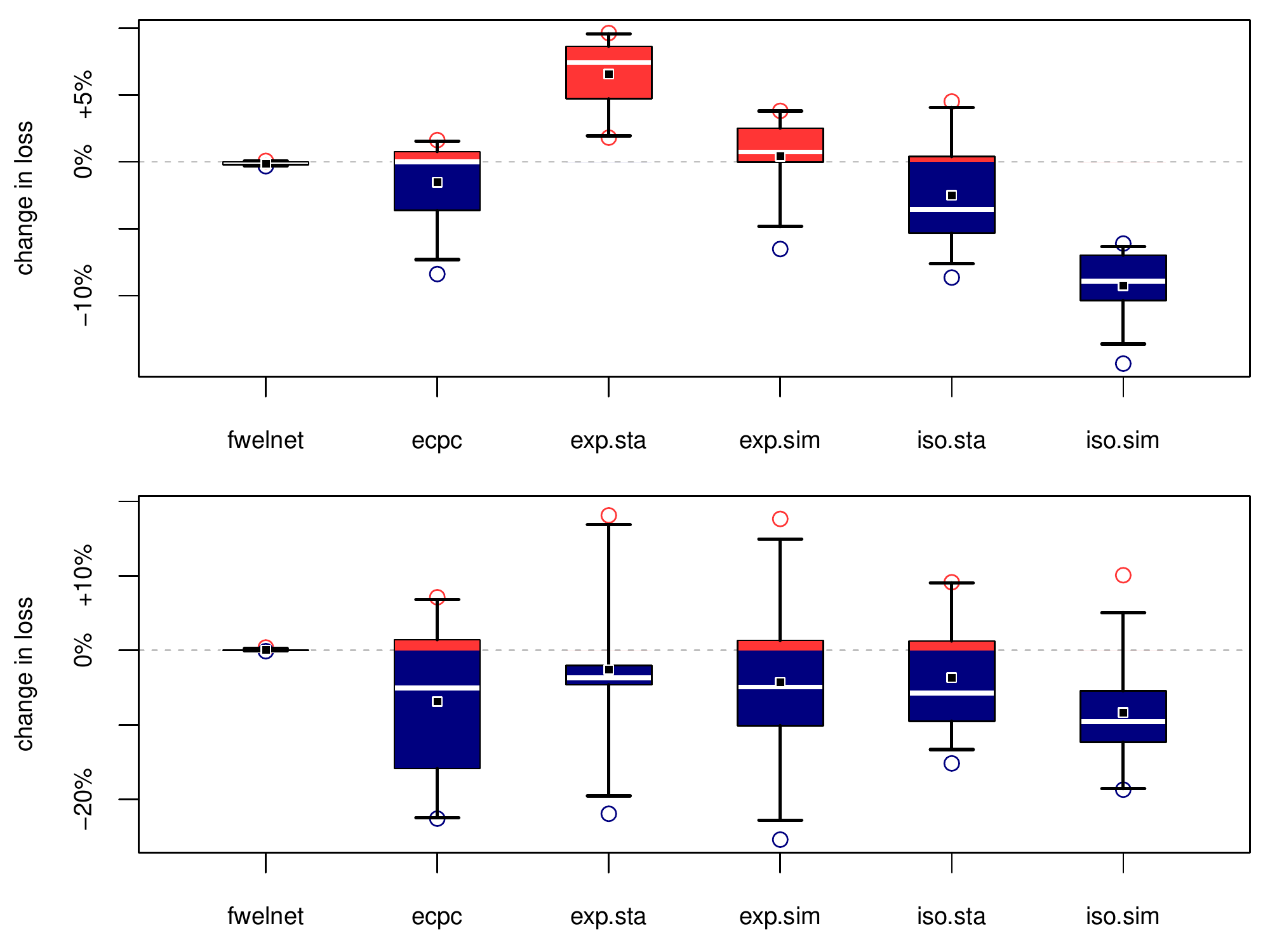}
\caption{Percentage change in cross-validated logistic deviance from ridge regression to other methods, for $10$ repetitions of $10$-fold cross-validation. Top: application on cervical cancer. Bottom: application on pre-eclampsia. From left to right: co-data learning (\texttt{fwelnet}, \href{https://CRAN.R-project.org/package=ecpc}{\texttt{ecpc}}), transfer learning with exponential/isotonic calibration and standard/simultaneous stacking. }
\end{figure}

\FloatBarrier

\hypertarget{internal-application}{%
\subsection{Internal application}\label{internal-application}}

In this application, we transfer information from a meta-analysis of
genome-wide association studies on Parkinson's disease
\citep[\textsc{pd-gwas},][]{Nalls2019} to the Luxembourg Parkinson's
study \citep[\textsc{l}ux\textsc{park},][]{Hipp2018}. The aim is to
classify samples into Parkinson's disease (\textsc{pd}) patients and
healthy controls based on single-nucleotide polymorphisms
(\textsc{snp}s).

At the time of our study, the \textsc{l}ux\textsc{park} data set
included genotyping and clinical data of \(790\) \textsc{pd} cases and
\(766\) healthy controls. \textsc{dna} samples were genotyped using the
NeuroChip array \citep{Blauwendraat2017}. Quality control steps of
genotyping data were conducted according to the standard procedures
reported previously \citep{Pavelka2022}. Missing genotyping data were
imputed using the reference panel from the Haplotype Reference
Consortium (release 1.1) on the Michigan Imputation Server
\citep{Das2016}
{[}\href{https://scicrunch.org/resolver/RRID:SCR_017579}{\textsc{rrid}:\textsc{id}\_017579}{]},
with a filter for imputation quality (\(r^2\) \textgreater{} 0.3).

As common \textsc{snp}s exhibit weak effects on \textsc{pd}, the sample
size is likely insufficient to train a highly predictive model. However,
publicly available summary statistics from the largest-to-date
\textsc{pd}-\textsc{gwas} (with around \(38\,000\) cases and
\(1\,400\,000\) controls from European ancestry) \citep{Nalls2019} might
serve as prior information on the \textsc{snp} effects. For each
\textsc{snp}, these summary statistics are the combined results from
simple logistic regression of the \textsc{pd} status on the
\textsc{snp}, namely the estimated slope (logarithmic odds ratio), its
standard error, and the associated \(p\)-value. Importantly, the
\textsc{l}ux\textsc{park} cohort was not part of the
\textsc{pd}-\textsc{gwas}, meaning that the prior information comes from
independent data. As the \textsc{l}ux\textsc{park} cohort and the
\textsc{pd-gwas} cohorts have a similar ethnic background, the prior
information might allow us to increase the predictive performance.

The two lists of \textsc{snp}s - from the \textsc{l}ux\textsc{park}
genotyping data (target data set) and the \textsc{pd-gwas} summary
statistics (source data set) - are partially overlapping. \textsc{snp}
data are high-dimensional and strongly correlated. From each block of
\textsc{snp}s in the target data set (250 kb window), we retain the most
significant one and those that are in weak pairwise linkage
disequilibrium with it (\(r^2 < 0.1\)). Next, we only retain the
\textsc{snp}s appearing also in the source data set. These two filtering
steps together reduce the dimensionality in the target data set from
around \(18\) million \textsc{snp}s to \(196\,018\) \textsc{snp}s. We
code the \textsc{snp} data for dominant effects, with \(0\) meaning no
alternate allele (\(0/0\)) and \(1\) meaning one or two alternate
alleles (\(0/1\) or \(1/1\)).

It seems that the results from the source data set are informative,
because \(5.80\%\) of the \(p\)-values are nominally significant at the
\(0.05\) level (\(11\,377\) out of \(196\,018\)), \(77\) are significant
at a false discovery rate of \(5\%\) (Benjamini-Hochberg), and \(35\)
are significant at a family-wise error rate of \(5\%\)
(Holm-Bonferroni). As \textsc{snp}s with a low minor allele frequency
might have large effect sizes but insignificant \(p\)-values, we base
the prior effects not on the estimated coefficients (\(\hat{\beta}\))
but on the signed logarithmic \(p\)-values
(\(-\mathrm{sign}(\hat{\beta})\log_{10}(p)\)). For each \textsc{snp}, we
compared the reference and the alternate alleles between the two data
sets: (i) If both data sets have the same reference allele and the same
alternate allele, the signed logarithmic \(p\)-value from the source
data set becomes the prior effect for the target data set. (ii) If the
reference allele of each data set is the alternate allele of the other
data set (swapped alleles), we invert the sign of the signed logarithmic
\(p\)-value. (iii) And if the two data sets have two different sets of
alleles (multiallelic \textsc{snp}), we set the prior effect to zero.

Rather than using the \(196\,018\) \textsc{snp}s for predictive
modelling in the target data set, we also filter them based on their
significance in the source data set (which is already a type of transfer
learning). For each cut-off in
\(\{5 \times 10^{-2},5 \times 10^{-3},\ldots,5 \times 10^{-10}\}\), we
exclude all \textsc{snp}s above and include all \textsc{snp}s below.
This means that for the target data set, we retain a specific number of
the most significant \textsc{snp}s from the source data set. For each
significance cut-off, we compare three modelling approaches:

\begin{itemize}

\item \textit{Uninformed approach}: We use logistic regression with ridge or lasso penalisation to model the \textsc{pd} status based on the included \textsc{snp}s. All included \textsc{snp}s are treated equally, irrespective of their estimated effect in the source data set.

\item \textit{Naïve transfer learning}: After calculating for each sample the sum across the signed logarithmic $p$-values from the source data set multiplied by the \textsc{snp}s from the target data set, we fit a simple logistic regression of the \textsc{pd} status on this sum.

\item \textit{Transfer learning}: The proposed transfer learning approach uses the signed logarithmic $p$-values from the source data set as prior effects for the target data set.

\end{itemize}

\begin{figure}
\centering
\includegraphics{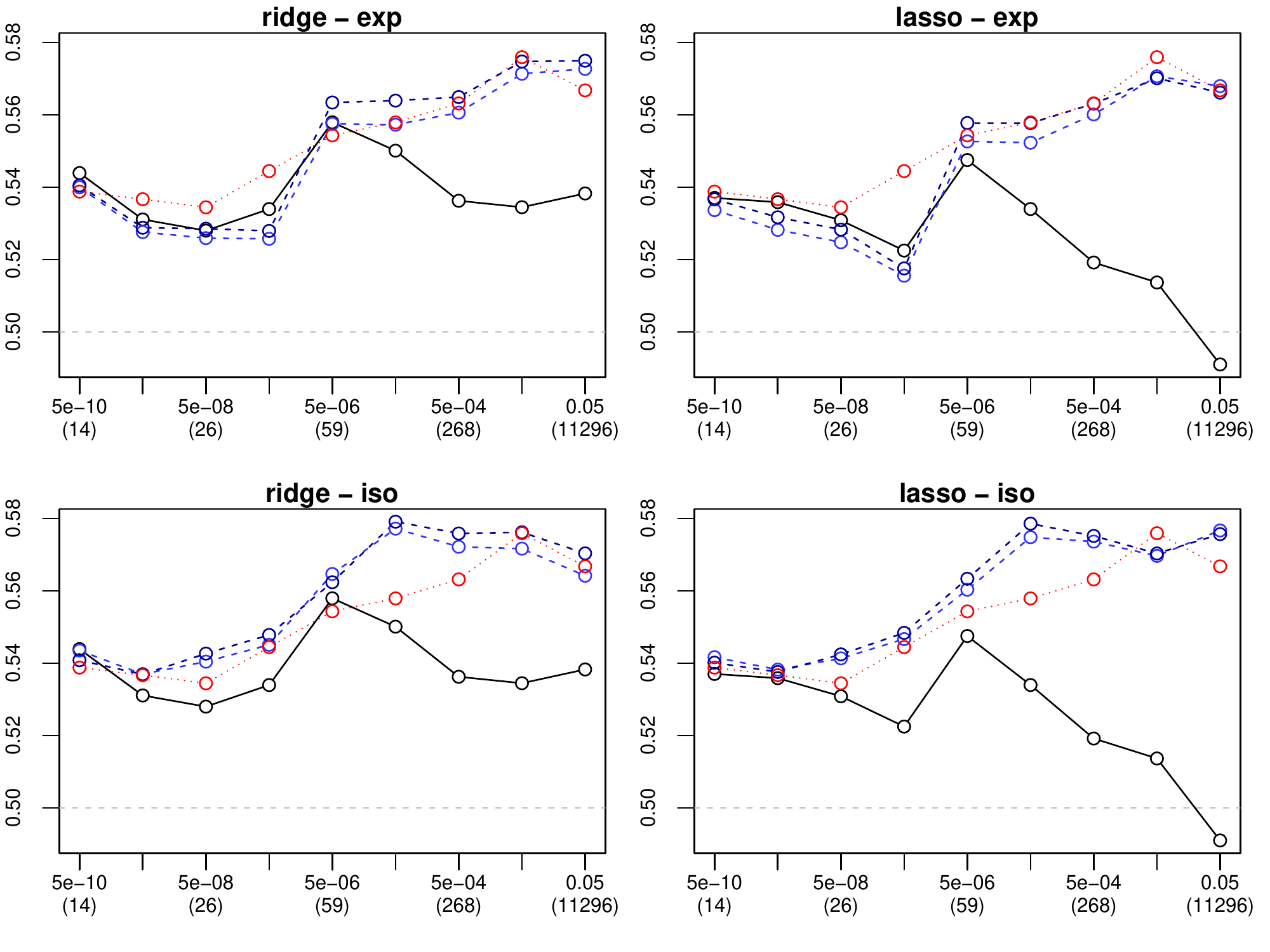}
\caption{Mean cross-validated concordance index (\textsc{c}-index/\textsc{auc}) from $10$ times $10$-fold cross-validation ($y$-axis) against $p$-value cutoff ($x$-axis) for regression without (solid line) and with (dashed lines) transfer learning (bright blue: standard stacking, dark blue: simultaneous stacking), under either ridge (left) or lasso (right) regularisation and either exponential (top) or isotonic (bottom) scaling. The numbers within brackets indicate the dimensionality and the dotted line is for naïve transfer learning.}
\label{tab_prs}
\end{figure}

Figure \ref{tab_prs} shows the predictive performance of modelling with
estimated effects (uninformed approach), with prior effects (naïve
transfer learning), or with both (transfer learning). We obtained the
results with repeated nested cross-validation (\(10\) repetitions,
\(10\) external folds, \(10\) internal folds), using the same folds for
all methods. If the significance cut-off is very strict, leading to a
small number of significant \textsc{snp}s, transfer learning does not
improve the predictive performance of ridge and lasso regression. In
these low-dimensional settings with many fewer \textsc{snp}s than
samples, prior information on the \textsc{snp}s is not helpful. But
otherwise transfer learning does improve the predictive performance of
ridge and lasso regression. This holds for all four flavours of the
proposed transfer learning method (exponential vs isotonic calibration,
standard vs simultaneous stacking), but isotonic calibration works
considerably better than exponential calibration and simultaneous
stacking works marginally better than standard stacking.

Depending on the significance cut-off determining the number of
significant \textsc{snp}s, the performance of naïve transfer learning
can be as high as the one of transfer learning with isotonic
calibration. In these cases, the prior effects are predictive to the
extent that it is not even necessary to estimate any effects. An
explanation for the high performance of naïve transfer learning might be
(i) the large sample size in the source data set for testing the
marginal effects of the \textsc{snp}s together with (ii) the linkage
disequilibrium clumping leading to a selection of relatively independent
\textsc{snp}s.

\FloatBarrier

\hypertarget{discussion}{%
\section{Discussion}\label{discussion}}

We proposed a two-step transfer learning method for exploiting estimated
coefficients from related studies to improve the predictive performance
in the study of interest. First, we adapt the prior effects from the
source data sets to the target data set, either with exponential or
isotonic calibration. While exponential calibration is more robust to
outliers (only three free parameters), isotonic calibration is more
flexible (only maintains order of prior effects). We expect the former
to be superior if the prior effects are close to the true effects, and
the latter to be superior if there is no exponential relationship.
Second, we combine the calibrated prior effects with information from
the observed data, based on two variants of stacked generalisation.
While the first variant (standard stacking) is more suitable if there
are many sources of co-data (\text{`}averaging calibrated prior effects
and estimated effects\text{'}), the second variant (simultaneous
stacking) is more suitable if there is one source of co-data with
partially unreliable or partially missing prior effects
(\text{`}shrinking combined effects towards calibrated prior
effects\text{'}).

The proposed transfer learning method allows for multiple sources of
prior information. It does not require the source data set(s) but only
the prior effects derived from the source data set(s). It therefore
allows researchers to transfer predictive information from one study to
another without requesting access to sensitive data. The proposed method
has a competitive predictive performance with an existing but less
flexible method (see simulation). In the case of closely related tasks,
accounting for prior effects with transfer learning seems to be more
beneficial than accounting for prior weights with co-data methods (see
application). We therefore believe that the proposed method could tackle
many biomedical predictions problems with one or more sets of prior
effects.

In some applications, only one type of prior information derived from
the source data sets is available. In other applications, multiple types
of prior information are available (or the source data sets themselves).
Then we can choose from multiple types of prior information. If the
source and target data sets have the same feature space, estimated
coefficients from penalised regression might be a reasonable choice. If
the feature spaces are different, however, it is problematic that (i)
lasso regression erratically selects among correlated features and (ii)
ridge regression distributes weight among correlated features. This
means that the presence or absence of additional correlated features in
the source data sets might change the prior information on the features
of interest. The same problem arises under contamination of a subset of
features \citep{Wiel2016}. We therefore expect that signed logarithmic
\(p\)-values (\(-\text{sign}(\hat{\beta}) \log_{10}(p)\)) will often be
more informative than estimated coefficients (\(\hat{\beta}\)).

\small

\hypertarget{funding}{%
\section{\texorpdfstring{\small Funding}{Funding}}\label{funding}}

This work was supported by the Luxembourg National Research Fund
(\textsc{fnr}) for the \textsc{era-n}et \textsc{eracos}y\textsc{sm}ed
\textsc{jtc}-2 project \textsc{pd-s}trat {[}\textsc{inter}/11651464{]}
and by the European Union's Horizon 2020 research and innovation
programme for the project \textsc{digi-pd} {[}\textsc{erapermed}
2020-314{]}. The National Centre of Excellence in Research on
Parkinson's Disease (\textsc{ncer-pd}) is funded by the \textsc{fnr}
{[}\textsc{fnr}/\textsc{ncer}13/\textsc{bm}/11264123{]}.

\hypertarget{acknowledgements}{%
\section{\texorpdfstring{\small Acknowledgements}{Acknowledgements}}\label{acknowledgements}}

We are grateful to Quentin \textsc{klopfenstein} for helpful discussions
on differential penalisation, and to the responsible and reproducible
research (R3) initiative for the pre-publication check. Data used in the
preparation of this manuscript were obtained from \textsc{ncer-pd}. All
participants provided written informed consent, and the study has been
approved by the National Ethics Board (\textsc{cner} Ref: 201407/13). We
would like to thank all participants of the Luxembourg Parkinson's Study
for their important support of our research. Furthermore, we acknowledge
the joint effort of the \textsc{ncer-pd} consortium members from the
partner institutions Luxembourg Centre for Systems Biomedicine,
Luxembourg Institute of Health, Centre Hospitalier de Luxembourg, and
Laboratoire National de Santé generally contributing to the Luxembourg
Parkinson's Study as listed below: Alexander \textsc{hundt}\(^{2}\),
Alexandre \textsc{bisdorff}\(^{5}\), Amir \textsc{sharify}\(^{2}\), Anne
\textsc{grünewald}\(^{1}\), Anne-Marie \textsc{hanff}\(^{2}\), Armin
\textsc{rauschenberger}\(^{1}\), Beatrice \textsc{nicolai}\(^{3}\), Brit
\textsc{mollenhauer}\(^{12}\), Camille \textsc{bellora}\(^{2}\), Carlos
\textsc{vega moreno}\(^{1}\), Chouaib \textsc{mediouni}\(^{2}\),
Christophe \textsc{trefois}\(^{1}\), Claire \textsc{pauly}\(^{1,3}\),
Clare \textsc{mackay}\(^{10}\), Clarissa \textsc{gomes}\(^{1}\), Daniela
\textsc{berg}\(^{11}\), Daniela \textsc{esteves}\(^{2}\), Deborah
\textsc{mcintyre}\(^{2}\), Dheeraj \textsc{reddy bobbili}\(^{1}\),
Eduardo \textsc{rosales}\(^{2}\), Ekaterina \textsc{soboleva}\(^{1}\),
Elisa \textsc{gómez de lope}\(^{1}\), Elodie \textsc{thiry}\(^{3}\),
Enrico \textsc{glaab}\(^{1}\), Estelle \textsc{henry}\(^{2}\), Estelle
\textsc{sandt}\(^{2}\), Evi \textsc{wollscheid-lengeling}\(^{1}\),
Francoise \textsc{meisch}\(^{1}\), Friedrich
\textsc{mühlschlegel}\(^{4}\), Gaël \textsc{hammot}\(^{2}\), Geeta
\textsc{acharya}\(^{2}\), Gelani \textsc{zelimkhanov}\(^{3}\), Gessica
\textsc{contesotto}\(^{2}\), Giuseppe \textsc{arena}\(^{1}\), Gloria
\textsc{aguayo}\(^{2}\), Guilherme \textsc{marques}\(^{2}\), Guy
\textsc{berchem}\(^{3}\), Guy \textsc{fagherazzi}\(^{2}\), Hermann
\textsc{thien}\(^{2}\), Ibrahim \textsc{boussaad}\(^{1}\), Inga
\textsc{liepelt}\(^{11}\), Isabel \textsc{rosety}\(^{1}\), Jacek
\textsc{jaroslaw lebioda}\(^{1}\), Jean-Edouard
\textsc{schweitzer}\(^{1}\), Jean-Paul \textsc{nicolay}\(^{19}\),
Jean-Yves \textsc{ferrand}\(^{2}\), Jens \textsc{schwamborn}\(^{1}\),
Jérôme \textsc{graas}\(^{2}\), Jessica \textsc{calmes}\(^{2}\), Jochen
\textsc{klucken}\(^{1,2,3}\), Johanna \textsc{trouet}\(^{2}\), Kate
\textsc{sokolowska}\(^{2}\), Kathrin \textsc{brockmann}\(^{11}\), Katrin
\textsc{marcus}\(^{13}\), Katy \textsc{beaumont}\(^{2}\), Kirsten
\textsc{rump}\(^{1}\), Laura \textsc{longhino}\(^{3}\), Laure
\textsc{pauly}\(^{1}\), Liliana \textsc{vilas boas}\(^{3}\), Linda
\textsc{hansen}\(^{1,3}\), Lorieza \textsc{castillo}\(^{2}\), Lukas
\textsc{pavelka}\(^{1,3}\), Magali \textsc{perquin}\(^{2}\), Maharshi
\textsc{vyas}\(^{1}\), Manon \textsc{gantenbein}\(^{2}\), Marek
\textsc{ostaszewski}\(^{1}\), Margaux \textsc{schmitt}\(^{2}\), Mariella
\textsc{graziano}\(^{17}\), Marijus \textsc{giraitis}\(^{2,3}\), Maura
\textsc{minelli}\(^{2}\), Maxime \textsc{hansen}\(^{1,3}\), Mesele
\textsc{valenti}\(^{2}\), Michael \textsc{heneka}\(^{1}\), Michael
\textsc{heymann}\(^{2}\), Michel \textsc{mittelbronn}\(^{1,4}\), Michel
\textsc{vaillant}\(^{2}\), Michele \textsc{bassis}\(^{1}\), Michele
\textsc{hu}\(^{8}\), Muhammad \textsc{ali}\(^{1}\), Myriam
\textsc{alexandre}\(^{2}\), Myriam \textsc{menster}\(^{2}\), Nadine
\textsc{jacoby}\(^{18}\), Nico \textsc{diederich}\(^{3}\), Olena
\textsc{tsurkalenko}\(^{2}\), Olivier \textsc{terwindt}\(^{1,3}\),
Patricia \textsc{martins conde}\(^{1}\), Patrick \textsc{may}\(^{1}\),
Paul \textsc{wilmes}\(^{1}\), Paula Cristina \textsc{lupu}\(^{2}\),
Pauline \textsc{lambert}\(^{2}\), Piotr \textsc{gawron}\(^{1}\), Quentin
\textsc{klopfenstein}\(^{1}\), Rajesh \textsc{rawal}\(^{1}\), Rebecca
\textsc{ting jiin loo}\(^{1}\), Regina \textsc{becker}\(^{1}\), Reinhard
\textsc{schneider}\(^{1}\), Rejko \textsc{krüger}\(^{1,2,3}\), Rene
\textsc{dondelinger}\(^{5}\), Richard \textsc{wade-martins}\(^{9}\),
Robert \textsc{liszka}\(^{14}\), Romain \textsc{nati}\(^{3}\), Rosalina
\textsc{ramos lima}\(^{2}\), Roseline \textsc{lentz}\(^{7}\), Rudi
\textsc{balling}\(^{1}\), Sabine \textsc{schmitz}\(^{1}\), Sarah
\textsc{nickels}\(^{1}\), Sascha \textsc{herzinger}\(^{1}\), Sinthuja
\textsc{pachchek}\(^{1}\), Soumyabrata \textsc{ghosh}\(^{1}\), Stefano
\textsc{sapienza}\(^{1}\), Sylvia \textsc{herbrink}\(^{6}\), Tainá
\textsc{marques}\(^{1}\), Thomas \textsc{gasser}\(^{11}\), Ulf
\textsc{nehrbass}\(^{2}\), Valentin \textsc{groues}\(^{1}\), Venkata
\textsc{satagopam}\(^{1}\), Victoria \textsc{lorentz}\(^{2}\), Walter
\textsc{maetzler}\(^{15}\), Wei \textsc{gu}\(^{1}\), Wim
\textsc{ammerlann}\(^{2}\), Yohan \textsc{jaroz}\(^{1}\), Zied
\textsc{landoulsi}\(^{1}\). \(^{1}\)Luxembourg Centre for Systems
Biomedicine, University of Luxembourg, Esch-sur-Alzette, Luxembourg,
\(^{2}\)Luxembourg Institute of Health, Strassen, Luxembourg,
\(^{3}\)Centre Hospitalier de Luxembourg, Strassen, Luxembourg,
\(^{4}\)Laboratoire National de Santé, Dudelange, Luxembourg,
\(^{5}\)Centre Hospitalier Emile Mayrisch, Esch-sur-Alzette, Luxembourg,
\(^{6}\)Centre Hospitalier du Nord, Ettelbrück, Luxembourg,
\(^{7}\)Parkinson Luxembourg Association, Leudelange, Luxembourg,
\(^{8}\)Oxford Parkinson's Disease Centre, Nuffield Department of
Clinical Neurosciences, University of Oxford, Oxford, UK, \(^{9}\)Oxford
Parkinson's Disease Centre, Department of Physiology, Anatomy and
Genetics, University of Oxford, Oxford, UK, \(^{10}\)Oxford Centre for
Human Brain Activity, Wellcome Centre for Integrative Neuroimaging,
Department of Psychiatry, University of Oxford, Oxford, UK,
\(^{11}\)Center of Neurology and Hertie Institute for Clinical Brain
Research, Department of Neurodegenerative Diseases, University Hospital
Tübingen, Tübingen, Germany, \(^{12}\)Paracelsus-Elena-Klinik, Kassel,
Germany, \(^{13}\)Ruhr-University of Bochum, Bochum, Germany,
\(^{14}\)Westpfalz-Klinikum GmbH, Kaiserslautern, Germany,
\(^{15}\)Department of Neurology, University Medical Center
Schleswig-Holstein, Kiel, Germany, \(^{16}\)Department of Neurology
Philipps, University Marburg, Marburg, Germany, \(^{17}\)Association of
Physiotherapists in Parkinson's Disease Europe, Esch-sur-Alzette,
Luxembourg, \(^{18}\)Private practice, Ettelbruck, Luxembourg,
\(^{19}\)Private practice, Luxembourg, Luxembourg.

\hypertarget{reproducibility}{%
\section{\texorpdfstring{\small Reproducibility}{Reproducibility}}\label{reproducibility}}

The \texttt{R} package \texttt{transreg} is available on \texttt{GitHub}
(\url{https://github.com/lcsb-bds/transreg}), with the code for the
simulations and the applications in a vignette
(\url{https://lcsb-bds.github.io/transreg/}). We obtained our results
using R 4.2.2
{[}\href{https://scicrunch.org/resolver/RRID:SCR_001905}{\textsc{rrid}:\textsc{id}\_001905}{]}
on a physical machine (aarch64-apple-darwin20, macOS Monterey 12.6).
Data for the application on cervical cancer are available from
\cite{Wiel2016}, in the \texttt{R} package
\href{https://doi.org/10.18129/B9.bioc.GRridge}{\texttt{GRridge}} in the
data set \text{`dataVerlaat'} (source data: \citealp{Farkas2013}, target
data: \citealp{Wiel2016}). Data for the application on pre-eclampsia are
available from \cite{Erez2017}, in the supporting file
\href{https://doi.org/10.1371/journal.pone.0181468.s001}{`pone.0181468.s001.csv'}.
For the application on Parkinson's disease, the source data are
available from \cite{Nalls2019}, in the online file
`nallsEtAl2019\_excluding23andMe\_allVariants.tab', and the target data
are available upon request
(\href{mailto:request.ncer-pd@uni.lu?cc=armin.rauschenberger@uni.lu,enrico.glaab@uni.lu&subject=data request Rauschenberger et al. ("transreg")}{\texttt{request.ncer-pd@uni.lu}}).
Information on reproducibility is also available on a frozen page (doi:
\href{https://doi.org/10.17881/hczj-3297}{10.17881/hczj-3297}).

\hypertarget{author-contributions}{%
\section{\texorpdfstring{\small Author
contributions}{Author contributions}}\label{author-contributions}}

\textsc{eg} acquired funding. \textsc{ar} and \textsc{m}vd\textsc{w}
developed the method. \textsc{ar} analysed the data and drafted the
manuscript. \textsc{zl} processed the \textsc{snp} data and critically
revised the internal application. \textsc{m}vd\textsc{w} and \textsc{eg}
critically revised the manuscript. All authors read and approved the
final manuscript.

\normalsize

\newpage

\hypertarget{appendix}{%
\section{Appendix}\label{appendix}}

\begin{table}[!hb]
\caption{Isotonic calibration. The aim is to (i) estimate the effects of the features under sign and order constraints determined by $q$ negative and $p-q$ non-negative prior effects, i.e. estimate  $\gamma_{1k},\ldots,\gamma_{pk}$ for $x_1,\ldots,x_p$ under $\hat{\gamma}_{jk}=0|z_{jk}=0$,
$\hat{\gamma}_{jk} \geq 0 | z_{jk} > 0$, $\hat{\gamma}_{jk} \leq 0 | z_{jk} < 0$, $\hat{\gamma}_{jk} \geq \hat{\gamma}_{lk} | z_{jk} \geq z_{lk}$, and $\hat{\gamma}_{jk} \leq \hat{\gamma}_{lk} | z_{jk} \leq z_{lk}$. This can be solved by (ii) estimating the effects of the features ordered by the co-data under sign and order constraints, i.e. estimate $\gamma_{(1),k},\ldots,\gamma_{(p),k}$ for $x_{(1)},\ldots,x_{(p)}$ under $\hat{\gamma}_{(j),k} \leq 0 | j \leq p$, $\hat{\gamma}_{(j),k} \geq 0 | j > p$, and $\hat{\gamma}_{(1),k} \leq \ldots \leq \hat{\gamma}_{(p),k}$. This in turn can be solved by (iii) estimating the effects of the combined features under sign constraints, i.e. estimate $\delta_{1},\ldots,\delta_{p}$ for $w_{1},\ldots,w_{p}$ under $\hat{\delta}_{j} \leq 0 | j \leq p$ and $\hat{\delta}_{j} \geq 0 | j > p$. Our algorithm receives the original features and the prior effects (i), orders the features by the prior effects (ii), combines the features (iii), estimates the effects of the combined features (iii), calculates the estimated effects of the ordered features (ii), and returns the estimated effects of the original features (i).}
\label{table_sub}
$$
\begin{array}{cccccc}
\hline & & & & & \\
\text{(i)} & x_{\circ,1}, z_{1,k} & x_{\circ,2}, z_{2,k} & \cdots & x_{\circ,q-1}, z_{q-1,k} & x_{\circ,q}, z_{q,k}
\\ & & & & & \\
\text{(ii)} & x_{\circ,(1)} & x_{\circ,(2)} & \cdots & x_{\circ,(q-1)} & x_{\circ,(q)}
\\ & & & & & \\
\text{(iii)} & \begin{array}{c} w_{\circ,1} = \\ x_{\circ,(1)} \end{array} & \begin{array}{c} w_{\circ,2} = \\ x_{\circ,(1)} + x_{\circ,(2)} \end{array} & \cdots & \begin{array}{c} w_{\circ,q-1} = \\ x_{\circ,(1)} + \cdots + x_{\circ,(q-1)} \end{array} & \begin{array}{c} w_{\circ,q} = \\ x_{\circ,(1)} + \cdots + x_{\circ,(q)} \end{array}
\\ & & & & & \\
\text{(iii)} & \hat{\delta}_{1,k} & \hat{\delta}_{2,k} & \cdots & \hat{\delta}_{q-1,k}  & \hat{\delta}_{q,k}
\\ & & & & & \\
\text{(ii)} & \begin{array}{c} \hat{\gamma}_{(1),k} = \\ \hat{\delta}_{1,k} + \cdots + \hat{\delta}_{q,k} \end{array} & \begin{array}{c} \hat{\gamma}_{(2),k} = \\ \hat{\delta}_{2,k} + \cdots + \hat{\delta}_{q,k} \end{array} & \cdots & \begin{array}{c} \hat{\gamma}_{(q-1),k} = \\ \hat{\delta}_{q-1,k} + \hat{\delta}_{q,k} \end{array} & \begin{array}{c} \hat{\gamma}_{(q),k} = \\ \hat{\delta}_{q,k} \end{array}
\\ & & & & & \\
\text{(i)} & \hat{\gamma}_{1,k} & \hat{\gamma}_{2,k} & \cdots & \hat{\gamma}_{q-1,k} & \hat{\gamma}_{q,k}
\\ & & & & & \\ \hline & & & & & \\
\text{(i)} & x_{\circ,q+1}, z_{q+1,k} & x_{\circ,q+2}, z_{q+2,k} & \cdots  & x_{\circ,p-1}, z_{p-1,k} & x_{\circ,p}, z_{p,k}
\\ & & & & & \\
\text{(ii)} & x_{\circ,(q+1)} & x_{\circ,(q+2)} & \cdots  & x_{\circ,(p-1)} & x_{\circ,(p)}
\\ & & & & & \\
\text{(iii)} & \begin{array}{c} w_{\circ,q+1} = \\ x_{\circ,(q+1)} + \cdots + x_{\circ,(p)} \end{array} & \begin{array}{c} w_{\circ,q+2} = \\ x_{\circ,(q+2)} + \ldots + x_{\circ,(p)} \end{array} & \cdots  & \begin{array}{c} w_{\circ,p-1} = \\ x_{\circ,(p-1)} + x_{\circ,(p)} \end{array} & \begin{array}{c} w_{\circ,p} = \\ x_{\circ,(p)} \end{array}
\\ & & & & & \\
\text{(iii)} & \hat{\delta}_{q+1,k} & \hat{\delta}_{q+2,k} & \cdots & \hat{\delta}_{p-1,k} & \hat{\delta}_{p,k}
\\ & & & & & \\
\text{(ii)} & \begin{array}{c} \hat{\gamma}_{(q+1),k} = \\ \hat{\delta}_{q+1,k} \end{array} & \begin{array}{c} \hat{\gamma}_{(q+2),k} = \\ \hat{\delta}_{q+1,k} + \hat{\delta}_{q+2,k} \end{array} & \cdots & \begin{array}{c} \hat{\gamma}_{(p-1),k} = \\ \hat{\delta}_{q+1,k} + \cdots + \hat{\delta}_{p-1,k} \end{array} & \begin{array}{c} \hat{\gamma}_{(p),k} = \\ \hat{\delta}_{q+1,k} + \cdots + \hat{\delta}_{p,k} \end{array}
\\ & & & & & \\
\text{(i)} & \hat{\gamma}_{q+1,k} & \hat{\gamma}_{q+2,k}  & \cdots & \hat{\gamma}_{p-1,k} & \hat{\gamma}_{p,k} \\
& & & & & \\ \hline
\end{array}
$$
\end{table}

\FloatBarrier

\newpage

\nocite{*}

\bibliographystyle{apalike}
\bibliography{bibliography}

\end{document}